\documentclass[cernpreprint, atlasdraft=false, texlive=2020, UKenglish, texmf]{atlasdoc}

\usepackage[subfigure=true]{atlaspackage}
\usepackage{atlasbiblatex}
\usepackage{adjustbox}
\usepackage{atlasphysics}
 
\graphicspath{{logos/}{figures/}}
 
\usepackage{ANA-FTAG-2020-08-PAPER-defs}

\AtlasTitle{Measurement of the $c$-jet mistagging efficiency in $t\bar{t}$~events
using $pp$ collision data at $\sqrt{s}=13$~$\text{Te\kern -0.1em V}$ collected with the ATLAS detector}
 
\PreprintIdNumber{CERN-EP-2021-166} 
 
\AtlasAbstract{ A technique is presented to measure the  efficiency
with which $c$-jets are mistagged as $b$-jets (mistagging efficiency)
using $t\bar{t}$ events, where one of the $W$
bosons decays into an electron or muon and a neutrino and the other
decays into a quark-antiquark pair. The measurement utilises the relatively large
and known $W\to cs$ branching ratio, which allows a measurement to be made
in an inclusive $c$-jet sample.  The data sample used was
collected by the ATLAS detector at $\sqrt{s} = 13$~$\text{Te\kern -0.1em V}$
and corresponds to an integrated luminosity of 139~fb$^{-1}$.
Events are reconstructed using a kinematic likelihood technique
which selects the mapping between jets and $t\bar{t}$
decay products that yields the highest likelihood value.
The distribution of the $b$-tagging discriminant for jets from the hadronic $W$
decays in data is compared with that in simulation to extract the mistagging efficiency
as a function of jet transverse momentum. The total uncertainties are in the range 3\%--17\%. The measurements generally agree with
those in simulation but there are some differences
in the region corresponding to the most stringent $b$-jet tagging requirement.

}
 
\ifdefined\INTNOTE
\AtlasAuthorContributor{Gwilliam, Carl}{a}{Setup AnalysisTop code and ran first set of ntuples}
\AtlasAuthorContributor{Li, Zhiyuan}{a}{High and low pt selection,  systematic errors, ntuple production.}
\AtlasAuthorContributor{Mehta, Andrew}{a}{Scale factor determination}
\AtlasAuthorContributor{Rompotis, Nikolaos}{a}{Event selection, systematic errors, ntuple production, pre and post-fit plotting}
\affil[a]{University of Liverpool}
\else
\author{The ATLAS Collaboration}
\fi
\AtlasRefCode{FTAG-2020-08}
 
\AtlasJournalRef{Eur. Phys. J. C 82 (2022) 95}
 
\AtlasDOI{DOI:10.1140/epjc/s10052-021-09843-w}

\hypersetup{pdftitle={ATLAS document},pdfauthor={The ATLAS Collaboration}}
 
\begin{document}
 
\maketitle

\section{Introduction}
\label{sec:intro}

The algorithms for the identification of jets containing $b$-hadrons,
also known as \btagging algorithms,
constitute an important tool for the analysis of the data collected by the ATLAS
experiment \cite{PERF-2007-01} at the Large Hadron Collider (LHC) \cite{Evans:2008zzb}.
Such algorithms play a crucial role in a large number of Standard Model (SM)
precision measurements (e.g.\ Refs.~\cite{TOPQ-2018-01,STDM-2017-38}),
Higgs boson measurements (e.g.\ Refs.~\cite{HIGG-2018-04,HIGG-2018-13}), and searches for supersymmetry
and other exotic phenomena (e.g.\ Refs.~\cite{EXOT-2019-13,SUSY-2018-38}).
 
The performance of the
\btagging algorithms depends crucially on whether the jets that are
identified contain $b$-hadrons (\bjets), $c$-hadrons (\cjets), or neither
of them (light-flavour jets).
The Monte Carlo (MC) simulation provides an estimate of the probabity that
a jet fulfils the requirements of the \btagging algorithm i.e.\ the
tagging efficiency  for \bjets\ and the mistagging
efficiency for \cjets\ and light-flavour jets. However, since the
simulation is not perfect, the (mis)tagging efficiency must be
measured in data. Simulation to data scale factors (SFs), which are defined as the ratio of
the efficiency measured in data to that in simulation, are used as a
correction to the simulation, assuming that the SFs
are independent of the physics process. Analyses that use
\btagging  apply a weight to each simulated event, derived as the
product of the SFs for all jets for which a tagging
requirement is made.
 
Measurements of the SFs have been made using \ttbar\
events for \bjets~\cite{PERF-2016-05,FTAG-2018-01} and inclusive jet
events for light-flavour jets~\cite{ATLAS-CONF-2018-006}. Measurements of
SFs for \cjets\ using proton--proton (\(pp\))
collision data at $\sqrt{s}= 7$~\TeV, described in Ref.~\cite{PERF-2012-04},
use two methods to measure the mistagging efficiency of \cjets. One
technique uses the production of a $W$ boson in association with a  \cjet\
that decays semi-muonically and the other identifies a $D^*$ meson within \cjets\ by explicitly
reconstructing the meson decay chain $D^{*+} \to D^{0} \pi^+ \to
K^{-}\pi^+\pi^+$.
 
This paper discusses a complementary technique to measure the inclusive
\cjet\ mistagging efficiency by means of a
\cjet\ sample that is derived from \ttbar\ events.  The measurement uses a
\(pp\) collision dataset which was collected with the ATLAS
detector during 2015--2018 and corresponds to an integrated luminosity of
\lumi.  A
likelihood-based method is used to select a high-purity \ttbar\
sample in which one of the $W$ bosons originating from the
top-quark decay $t\to W b$ decays leptonically
into an electron or a muon plus the
corresponding neutrino and the other decays hadronically. The
branching ratio of a $W$ boson to
final states containing a charm-quark, which is approximately
33\%~\cite{Zyla:2020zbs},
allows the \cjet\ mistagging efficiency of the data to be
determined as a function of jet transverse momentum by fitting
simulated event samples to the data sample.
The technique has the advantage that it does not depend on a specific
hadron decay chain topology.

The measurement of the \cjet\ mistagging efficiency presented in this paper uses particle-flow (PFlow) jets, which are made by combining
calorimeter energy deposits with inner-detector tracks~\cite{PERF-2015-09},
but the same technique is also used to make measurements for
jets made only from calorimeter energy deposits~\cite{JETM-2018-05}
and  jets that only use tracks (track-jets)~\cite{ATL-PHYS-PUB-2017-010}.
 
This paper is structured as follows. Section~\ref{sec:atlas} describes
the ATLAS detector. Data and simulated samples are discussed in
Section~\ref{sec:samples}.  The reconstruction of electrons,
muons, jets and missing transverse momentum is described in
Section~\ref{sec:objects}, and the \btagging algorithms are discussed
in Section~\ref{sec:btagging}.  The selection of the \ttbar\ sample is
described in Section~\ref{sec:eventsel}. The measurement of the
\cjet\ mistagging efficiency is described in  Section~\ref{sec:analysis},
and Section~\ref{sec:conclusion} gives the conclusions.
 
\section{ATLAS detector}
\label{sec:atlas}
 
The ATLAS detector~\cite{PERF-2007-01} is a multipurpose particle physics detector with a forward--backward symmetric cylindrical
geometry and nearly 4$\pi$ coverage in solid angle.\footnote{ATLAS uses a right-handed coordinate system with its origin at the nominal
interaction point in the centre of the detector. The positive $x$-axis is defined by the direction from the interaction point to the centre
of the LHC ring, with the positive $y$-axis pointing upwards, while the beam direction defines the $z$-axis. Cylindrical coordinates
$(r,\phi)$ are used in the transverse plane, $\phi$ being the azimuthal angle around the $z$-axis. The component of momentum
in the transverse plane is denoted by \pt. The pseudorapidity $\eta$ is
defined in terms of the polar angle $\theta$ by $\eta=-\ln\tan(\theta/2)$. Rapidity is defined as $y$ = $0.5 \ln[(E + p_z)/(E-p_z)]$
where $E$ denotes the energy, and $p_z$ is the component of the momentum along the beam direction.
The separation of two objects in $\eta$--$\phi$ space is given by $\Delta R=\sqrt{(\Delta \eta)^2 + (\Delta \phi)^2}$.}
The inner tracking detector consists of silicon pixel and  microstrip detectors
covering the pseudorapidity region $|\eta| < 2.5$, surrounded by a transition radiation tracker
which enhances electron identification in the region $|\eta| < 2.0$.
Between Run~1 and Run~2, a new inner pixel layer, the insertable B-layer~\cite{ATLAS-TDR-2010-19, PIX-2018-001}, was added at a mean sensor radius of
3.3~cm.
The inner detector is surrounded by a thin superconducting solenoid providing an axial 2~T magnetic field, and by a fine-granularity
lead/liquid-argon (LAr) electromagnetic calorimeter covering $|\eta| < 3.2$. A steel/scintillator-tile calorimeter provides hadronic coverage
in the central pseudorapidity range ($|\eta| < 1.7$).
The endcap and forward regions ($1.5 < |\eta| < 4.9$) of the hadronic calorimeter are made of LAr active layers with either copper or tungsten
as the absorber material.
An extensive muon spectrometer (MS) with an air-core toroidal magnet system surrounds the calorimeters.
Three layers of high-precision tracking chambers provide coverage in the range $|\eta| < 2.7$, while
dedicated fast chambers allow triggering in the region $|\eta| < 2.4$.
The ATLAS trigger system consists of a hardware-based level-1 trigger followed by a software-based high-level trigger~\cite{TRIG-2016-01}.
An extensive software suite~\cite{ATL-SOFT-PUB-2021-001} is used in the reconstruction
and analysis of real and simulated data, in detector operations, and in the trigger and data acquisition systems of the experiment.

\section{Data and simulated event samples}
\label{sec:samples}
The data analysed in this paper correspond to
139~fb$^{-1}$~\cite{ATLAS-CONF-2019-021,LUCID2}  of
\(pp\) collision data collected by the ATLAS detector between 2015 and
2018 with a centre-of-mass energy of 13~\TeV\ and a 25\,ns proton
bunch crossing interval.  The data sample was collected using a set of
single-electron~\cite{TRIG-2018-05} and single-muon~\cite{TRIG-2018-01}
triggers  with \pt\ thresholds in the range of 20--26~\GeV\  depending on the
lepton flavour and  data-taking period.
All detector subsystems were required to be operational
during data taking and to fulfil data quality requirements. Events with noise bursts or coherent
noise in the calorimeters are removed.  The
presence of additional interactions in the same bunch crossing,
referred to as pile-up, is characterised by the average number of such
interactions, \averagemu, which was 34 for the whole dataset.

Simulated event samples are used to model SM
processes and to estimate the expected signal yields.  All samples were
produced using the ATLAS simulation infrastructure~\cite{SOFT-2010-01}
and $\GEANT4$~\cite{Agostinelli:2002hh}. A subset of samples used a faster
simulation based on a parameterisation for the calorimeter response and
$\GEANT4$ for the other detector systems~\cite{SOFT-2010-01}.
The simulated events are reconstructed with the same algorithms as
used for data, and contain a realistic modelling of pile-up
interactions. The pile-up profiles in the simulation match those of each dataset
between 2015 and 2018, and were obtained by overlaying the hard-scatter events
with minimum-bias events simulated using the soft QCD processes of
{\PYTHIA}\,8~\cite{Sjostrand:2014zea} with
the \nnpdflo set~\cite{Ball:2014uwa} of
parton distribution functions (PDFs)~\cite{Ball:2012cx} and
a set of tuned parameters called the A3 tune~\cite{ATL-PHYS-PUB-2016-017}.
For all samples, with the exception of those generated using
\sherpa~\cite{Bothmann:2019yzt}, the decays of bottom and charm
hadrons were performed by \evtgen~\cite{Lange:2001uf}.
 
The events used in this study originate mostly
from  \ttbar\ production. This process was modelled using the
{\powhegbox}\,v2~\cite{Frixione:2007nw,Nason:2004rx,Frixione:2007vw,Alioli:2010xd}
generator at next-to-leading order (NLO)
with the \nnpdfnlo PDF set~\cite{Ball:2014uwa}
and the \hdamp\ parameter\footnote{The
\hdamp\ parameter is a resummation damping factor and one of the
parameters that controls the matching of \powheg matrix elements to
the parton shower and thus effectively regulates the
high-\pt\ radiation against which the \ttbar\ system recoils.} set
to $1.5 \times \mtop$~\cite{ATL-PHYS-PUB-2016-020},
where \mtop\ denotes the mass of the top quark.  The events were interfaced
to {\PYTHIA}\,8.230 to model the parton shower,
hadronisation, and underlying event, with parameters set according
to the A14 tune~\cite{ATL-PHYS-PUB-2017-007} and using the \nnpdftwo set of PDFs~\cite{Ball:2014uwa}.
The uncertainty due to initial-state radiation (ISR) is estimated by
simultaneous variations of the \hdamp\ parameter and the renormalisation and
factorisation scales, and choosing the Var3c up/down variants of the A14
tune.  The impact of final-state
radiation (FSR) is evaluated by varying the renormalisation scale for
emissions from the parton shower by a factor two up or down.
The impact of using a different parton shower and
hadronisation model is evaluated by comparing the nominal \ttbar\ sample
with another \ttbar\ sample generated by {\powhegbox}\,v2 using exactly the same
parameters but interfaced to {\herwig}~7.04~\cite{Bahr:2008pv,Bellm:2015jjp},
using the H7UE tune~\cite{Bellm:2015jjp}
and the \mmhtlo~PDF set~\cite{Harland-Lang:2014zoa}.
 
In addition to \ttbar\ production, some minor backgrounds
contribute to the final event sample used for the calibration.
These backgrounds consist mostly of single-top and diboson production,
the production of \ttbar\ in association with a vector boson
and the production of a vector boson in association with jets.
Details of the modelling of these samples are given in the following.
 
Single-top $s$-channel production was modelled using the {\powhegbox}\,v2
generator at NLO in QCD in the five-flavour scheme with the \nnpdfnlo PDF set.
Single-top $t$-channel production was modelled using the
{\powhegbox}\,v2
generator at NLO in QCD, using the four-flavour scheme and the
corresponding \nnpdfnlo set of PDFs.
The associated production of a single top quark and a $W$ boson ($tW$) was
modelled using the
{\powhegbox}\,v2
generator at NLO in QCD, using the five-flavour scheme and the
\nnpdfnlo set of PDFs.
The diagram removal scheme~\cite{Frixione:2008yi} was used to
remove interference and overlap with \ttbar\ production.
The events for all single-top production channels
were interfaced to {\PYTHIA}\,8.230
using the A14 tune
and the \nnpdftwo set of PDFs.
 
The production of $Z+$jets and $W$+jets events was simulated with the
{\sherpa}\,2.2.1
generator using NLO matrix elements for up to two partons, and leading-order (LO) matrix elements
for up to four partons calculated with the Comix~\cite{Gleisberg:2008fv}
and \openloops~\cite{Buccioni:2019sur,Cascioli:2011va,Denner:2016kdg} libraries. They
were matched with the \sherpa parton shower~\cite{Schumann:2007mg} using the MEPS@NLO
prescription~\cite{Hoeche:2011fd,Hoeche:2012yf,Catani:2001cc,Hoeche:2009rj}
and the set of tuned parameters developed by the \sherpa authors.
The \nnpdfnnlo set of PDFs~\cite{Ball:2014uwa} was used and the samples
were normalised to a next-to-next-to-leading-order (NNLO)
prediction~\cite{Anastasiou:2003ds}.
 
Samples of events with diboson final states ($VV$) were simulated with the
{\sherpa}\,2.2.1 or 2.2.2
generator depending on the process,
including off-shell effects and Higgs boson contributions where appropriate.
Fully leptonic final states and semileptonic final states, where one boson
decays leptonically and the other hadronically, were generated using
matrix elements at NLO accuracy in QCD for up to one additional parton
and at LO accuracy for up to three additional parton
emissions. Event samples for the loop-induced processes $gg \to VV$ were
generated using LO-accurate matrix elements for up to one
additional parton emission for both the fully leptonic and
semileptonic final states. The matrix element calculations were matched
and merged with the \sherpa parton shower based on Catani--Seymour
dipole factorisation~\cite{Gleisberg:2008fv,Schumann:2007mg} using the MEPS@NLO
prescription.
The virtual QCD corrections were provided by the
\openloops library. The
\nnpdfnnlo set of PDFs was used,
along with the dedicated set of tuned parton-shower parameters developed by the
\sherpa authors.
 
Production of \ttbar\ in association with a vector boson
was modelled using the
{\mgamc}\,2.3.3 \cite{Alwall:2014hca} generator at NLO with the
\nnpdfnlo PDF set.
The events were interfaced to {\PYTHIA}\,8.210~\cite{Sjostrand:2014zea}
using the A14 tune and the \nnpdftwo PDF set.

\section{Object reconstruction}
\label{sec:objects}
 
Selected events are required to contain at least one vertex having at least two associated tracks with $\pT > 500$~\MeV, and the primary vertex is chosen to be the vertex reconstructed with the largest $\Sigma p_{\mathrm{T}}^2$ of its associated tracks.
 
Electron candidates are reconstructed by matching inner-detector
tracks to clusters of energy deposited in the EM calorimeter.
Electrons must have $\pt>27$~\GeV\ and $|\eta|<2.47$. The
associated track must have $|d_0|/\sigma_{d_0}<5$ and $|z_0|\sin
\theta<0.5$~mm, where $d_0$ ($z_0$) is the transverse (longitudinal)
impact parameter relative to the primary vertex and $\sigma_{d_0}$ is
the uncertainty in $d_0$. Candidates are identified with a
likelihood method and must satisfy the `medium' identification
criteria described in Ref.~\cite{EGAM-2018-01}. The likelihood relies
on the shape of the EM shower measured in the calorimeter, the quality
of the track reconstruction, and the quality of the match between the
track and the cluster. To suppress candidates originating from photon
conversions, hadron decays, or jets misidentified as electrons,
candidates are required to satisfy the gradient isolation criteria
based on tracking and calorimeter measurements~\cite{EGAM-2018-01}.
The electron energy and reconstruction efficiency are calibrated using $Z \to e^+e^-$
decays~\cite{EGAM-2018-01}.
 
Muon candidates are reconstructed in the range $|\eta|<2.5$ by
combining tracks in the inner detector with tracks in the MS.
All muon candidates must have $\pt>27$~\GeV,
$|d_0|/\sigma_{d_0}<3$, and $|z_0|\sin \theta<0.5$~mm.  The `medium' quality
requirements described in Ref.~\cite{ATLASMuonpub} are used
and muons from hadron decays are suppressed by imposing a track-based isolation
requirement. The muon
reconstruction efficiency in
the simulation is corrected using comparisons with data~\cite{MUON-2018-03}.
 
Jets are formed using objects from a
particle-flow  algorithm, which combines  energy
deposits in the calorimeter with inner detector
tracks~\cite{PERF-2015-09}. The PFlow objects are combined into
jets  in the
range $|\eta|<2.5$ and $\pt>20$~\GeV\ using the anti-$k_t$
algorithm~\cite{Cacciari:2008gp,Fastjet} with a radius parameter $R$ of
$0.4$. A jet-vertex-tagging
technique using a multivariate likelihood~\cite{PERF-2014-03} is
applied to jets with $|\eta|<2.4$ and $\pt<60$~\GeV\ to suppress jets
that are not associated with the event's primary vertex.
Jets are
further calibrated according to in situ measurements of the jet energy
scale~\cite{JETM-2018-05}.
 
The labelling
scheme used to define the flavour of a jet in a simulated event is applied by
matching reconstructed jets to  generator-level $b$- or $c$-hadrons
with $\pt > 5$~\GeV\ within a cone of size $\Delta R = 0.3$ around the
jet axis. Jets that contain a $b$-hadron are called \bjets.
Remaining jets containing a $c$-hadron are called \cjets. Jets with a
hadronically decaying $\tau$-lepton are called $\tau$-jets, and all remaining
jets are called light-flavour jets.
 
Overlaps between reconstructed objects are removed using a procedure
based on the angular separation between different final-state objects.
The procedure is similar to the one described in Ref.~\cite{TOPQ-2018-15}.

The event's missing transverse momentum, whose magnitude is denoted by
\MET, is computed as the negative vectorial sum of the transverse
momenta of leptons,
jets
and a track-based soft term~\cite{PERF-2016-07} accounting for the
contribution from particles from the primary vertex that are not
already included. The jets employed in the \MET\ calculation
include PFlow jets and, in addition,
anti-$k_t$ $R = 0.4$ jets with $\pt>30$~\GeV\ and
$2.5<|\eta|<4.5$.

\section{Description of \btagging algorithms  }
\label{sec:btagging}
Jets that contain a $b$-hadron are distinguished from other jets that
contain a $c$-hadron or only light-flavour hadrons mainly by the
larger mass and longer lifetime of the $b$-hadron.  The \btagging
algorithm studied in this paper is called DL1r, and is an updated
version of the DL1 tagger described in Ref.~\cite{ATL-PHYS-PUB-2017-013}.
Information from the tracks in the jet, such as their transverse impact parameter
and the reconstructed secondary and tertiary vertices, are combined
into a set of low-level taggers. The DL1 algorithm combines the output of the
low-level taggers into a single discriminant by using a deep neural network.
DL1r adds the result of the RNNIP algorithm, which is
based on a recurrent neural network exploiting the correlation between
the tracks' impact parameters.
Current analyses in ATLAS use the DL1r
discriminant in intervals defined by the efficiency to tag \bjets\ in
simulated $t\bar{t}$ events~\cite{FTAG-2018-01}.  The  \btagging interval
boundaries are 0\%, 60\%, 70\%, 77\%, 85\% and 100\%.
 
Measurements of the  tagging efficiency SFs for \bjets\ are made
as a function of jet \pt\ for $|\eta|<2.5$,
using $t\bar{t}$ events where both top
quarks decay leptonically and a method similar to that described
in Ref.~\cite{FTAG-2018-01}. The SFs for the mistagging efficiency
for light-flavour jets are obtained by defining a `flipped tagger'
in which the sign of the track
impact parameters that are used in the low-level
taggers is inverted~\cite{ATLAS-CONF-2018-006}. This results in similar
efficiencies for light- and heavy-flavour jets, allowing the
light-flavour SFs to be determined. Both the $b$- and
light-flavour jet SFs have been determined using the full Run~2 dataset. It
should be noted that both sets of SFs require a
determination of the \cjet\ SFs, which were taken from a
preliminary version of the method described in this paper. Since
the  contamination from \cjets\ is relatively small in these measurements
and because the previous SFs were similar to those in the present measurement, any differences only have a small impact.

Analyses that use \btagging in ATLAS apply a weight to each simulated event,
derived as the product of the SFs for all jets for which a
tagging requirement is made. It is found that there are significant
differences between the simulated (mis)tagging efficiency for different
fragmentation models~\cite{FTAG-2018-01}. To account for these
differences, simulation-to-simulation SFs are applied to
those simulated samples that have a fragmentation model and decay different
from the one used to
derive the SFs. In the present paper the
$t\bar{t}$ simulation (see Section~\ref{sec:samples}) uses
{\PYTHIA}\,8 and \evtgen, which is the reference MC fragmentation program for all the measured SFs.

\section{Event selection and reconstruction}
\label{sec:eventsel}
The analysis aims to obtain a pure sample of semileptonic \ttbar\
events.  Exactly one electron or one muon, denoted by $\ell$, is
required.  The
events are also required to have $\MET > 20$~\GeV, and the transverse
mass $\mt$ of the \MET\ and the lepton must satisfy
\[ \mt = \sqrt{2 \pt^\ell \MET (1-\cos\Delta\phi)} > 40~\GeV,\]
where $\Delta\phi = \phi(\MET)-\phi(\ell)$ is the azimuthal angle between
the lepton and \MET. Subsequently, a
requirement on the number of PFlow jets is applied such that one of the
following selections is true:
\begin{itemize}
\item the event contains exactly four jets with $\pt > 25$~\GeV\
\item the event contains at least three jets with $\pt > 25$~\GeV\
and exactly one jet with $20 < \pt < 25$~\GeV
\item the event contains at least five jets with $\pt > 25$~\GeV\
and at least one of these jets has  $\pt > 70$~\GeV.
\end{itemize}
These selections are designed to keep a high number of jets covering a
large range in $\pt$, while reducing the non-$t\bar{t}$ background and
the number of jets arising from final state QCD radiation (FSR).  The first
requirement is the `baseline' selection, which has a relatively low
rate of FSR jets, due to requiring exactly four jets.  The second
requirement allows measurements to be made for jets bewteen $20 < \pt
< 25$~\GeV, while minimising the non-$t\bar{t}$ background which is
more likely to have multiple low $\pt$ jets. The third requirement
improves the statistics of high $\pt$ jets where the rate of FSR is greater.
Although allowing five jets increases the fraction FSR jets  being included in the selection,  the rate of such jets is reasonable for jet
$\pt > 70$~\GeV, since the likelihood, described below, is better able to identify the top decay products at higher \pt.
 
The four-vectors of the four highest-\pt\ jets, the lepton and the
event \MET\ are used as inputs to a likelihood-based \ttbar\ event
reconstruction algorithm, which minimises deviations in the invariant
masses of the top quarks and $W$ bosons from their true values
and is described in more detail in
Ref.~\cite{ERDMANN201418}. This algorithm uses a likelihood function
to assign the four jets to the \ttbar\ decay topology. In particular,
the algorithm assigns one jet to the \bjet\ from the leptonically
decaying top quark ($t\to Wb \to \ell \nu b$), another jet to  the \bjet\
from the hadronically decaying top quark ($t\to Wb \to qq^\prime b$,
where $qq^\prime$ are the quarks into which the $W$ boson decays) and
the remaining two jets to the jets that come from the hadronic $W$
boson decay. The jet assignment does not use any \btagging information.
The following notation is used: the jets that are
assigned as the decay products of the $W$ boson are referred to as
$W$-jets and the remaining two jets are referred to as top-jets.

The four-jet combination that has the highest likelihood is selected
for the \ttbar\ reconstruction only if the negative logarithm of the
likelihood value is greater than $-48$. This requirement increases the fraction
of events where the top-quark decay products are correctly assigned.
Subsequently, the top-jets are required to lie in the tightest
(0--60\%) DL1r \btagging interval, whereas no \btagging requirement is
imposed on the $W$-jets, so as to leave them unbiased. After these
requirements it is found that 98.8\% of events in the $t\bar{t}$ simulation
have two top-jets that are \bjets.
 
\begin{figure}
\centering
\includegraphics[width=0.45\textwidth]{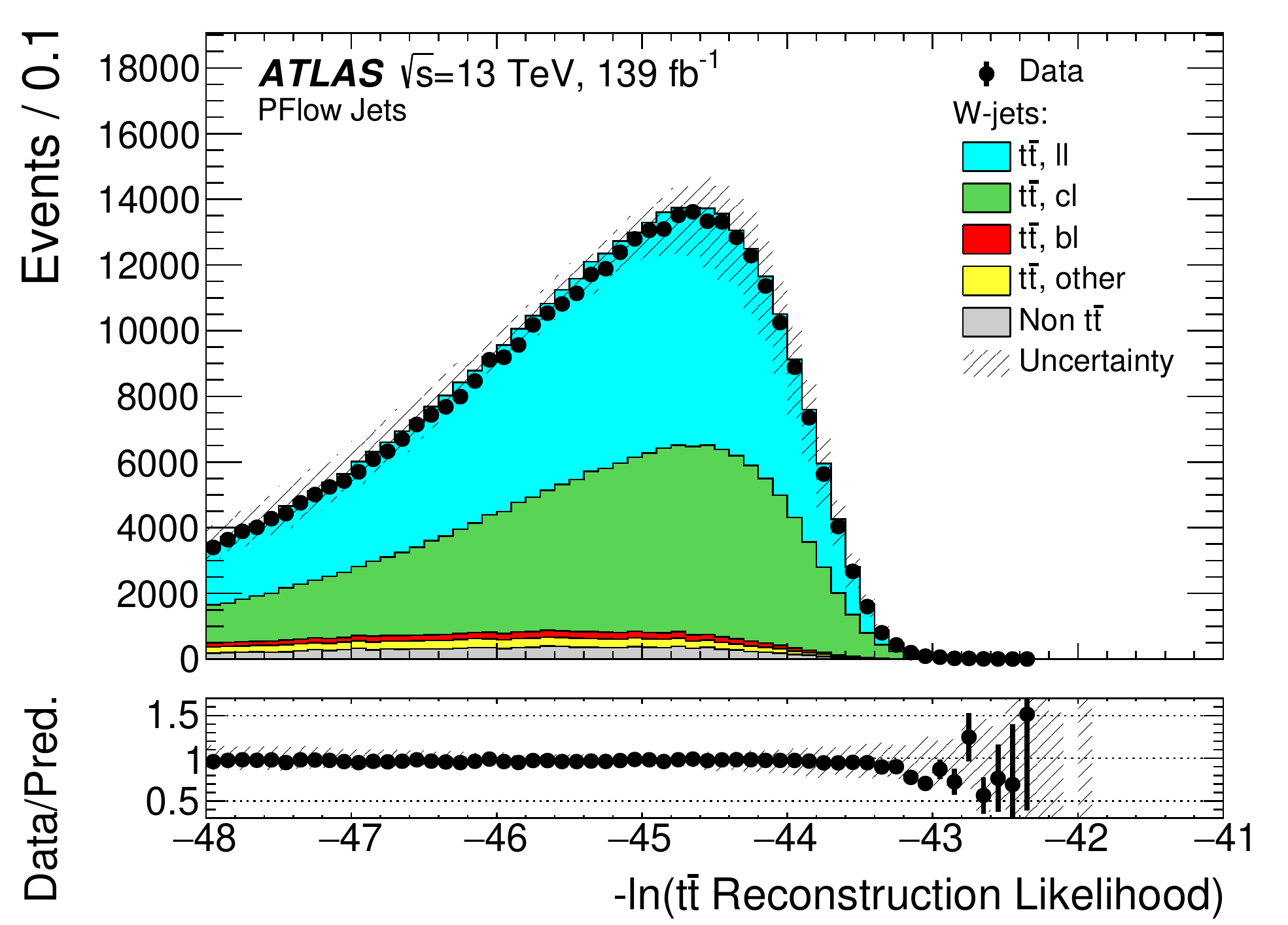}
\caption{Distribution of  the negative logarithm of the likelihood that
is used to reconstruct the \ttbar\ decay.
Only events with values of this quantity exceeding -48 are shown,
which corresponds to the selection used in the measurement.
The $t\bar{t}$ simulation is split according
to the flavours of the $W$-jets ($ll$, $cl$, $bl$ and other). The (mis)tagging efficiency scale factors have not been applied to the simulation. The hashed area shows the total uncertainty, excluding the uncertainties from the (mis)tagging efficiency scale factors
and the uncertainty on the jet $\pt$ distribution, which is derived from the difference between data and simulation.}
\label{fig:llr}
\end{figure}

The distribution of
the value of the negative logarithm of the likelihood for
the selected events is shown in Figure~\ref{fig:llr}.
The majority of the events come from \ttbar\ production with 3\%
of the events, referred to as `non \ttbar', coming from other processes such as $W$+jets or $Z$+jets production
or single-top production. The simulated \ttbar\
events are categorised according to the flavour of the $W$-jets. The notation
`$ll$' indicates that both $W$-jets are light-flavour jets.
Similarly, `$cl$' (`$bl$')
indicates that one of the $W$-jets is a \cjet\ (\bjet)
and the other is a light-flavour jet. The $W$-jet pairs with flavours
other than those discussed above fall into a
category called `other', which also  includes
events in which at least one of the $W$-jets comes from a
hadronically decaying $\tau$-lepton. The $ll$ category accounts
for 55\% of events, the $cl$ category for $41$\%, the $bl$ category for 1.8\%
and the `other' category for 2.5\%.
 
The simulated \ttbar\ yield shown in Figure~\ref{fig:llr}, as well as
in all subsequent figures and results in this paper, is corrected
for compatibility with ATLAS \ttbar\ measurements~\cite{TOPQ-2017-12},
which indicate that the simulation
underestimates the number of events containing more than two
\bjets.
For this reason, simulated events
in which both the top-jets and at least one of the $W$-jets are \bjets\ are
scaled by $1.25 \pm 0.25$, where the full difference between data and
the prediction from simulation is taken as a systematic uncertainty. The remaining events are normalised using the prediction from simulation. Finally,
Figure~\ref{fig:llr} includes an uncertainty in the simulation estimate (see Section~\ref{sec:systematics}),
derived as the combination of the detector systematic uncertainties, the \ttbar\
modelling uncertainties  and the
uncertainty due to the limited number of simulated events.
 
\begin{figure}
\centering
\subfigure[]{\includegraphics[width=0.45\textwidth]{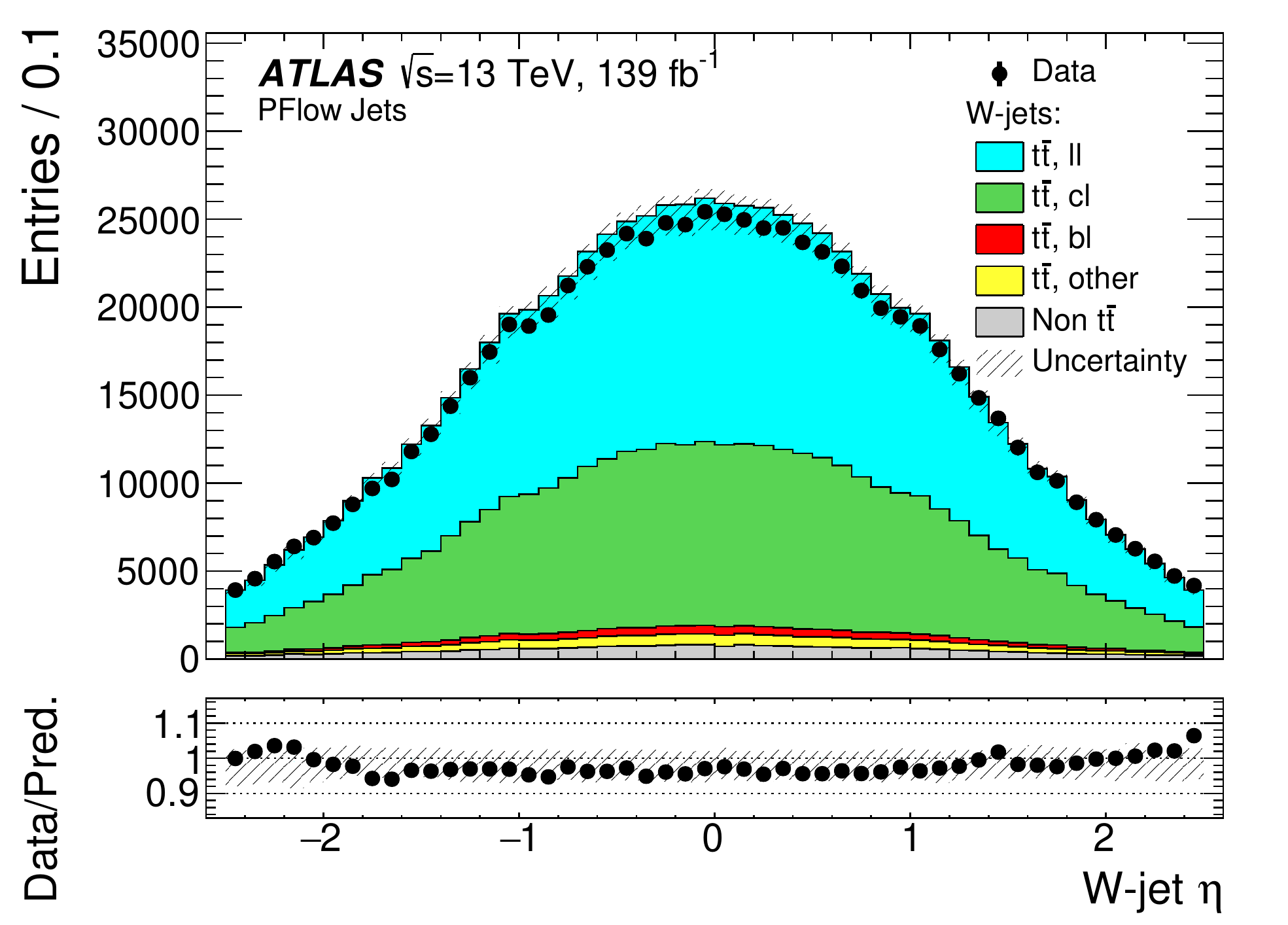}}
\subfigure[]{\includegraphics[width=0.45\textwidth]{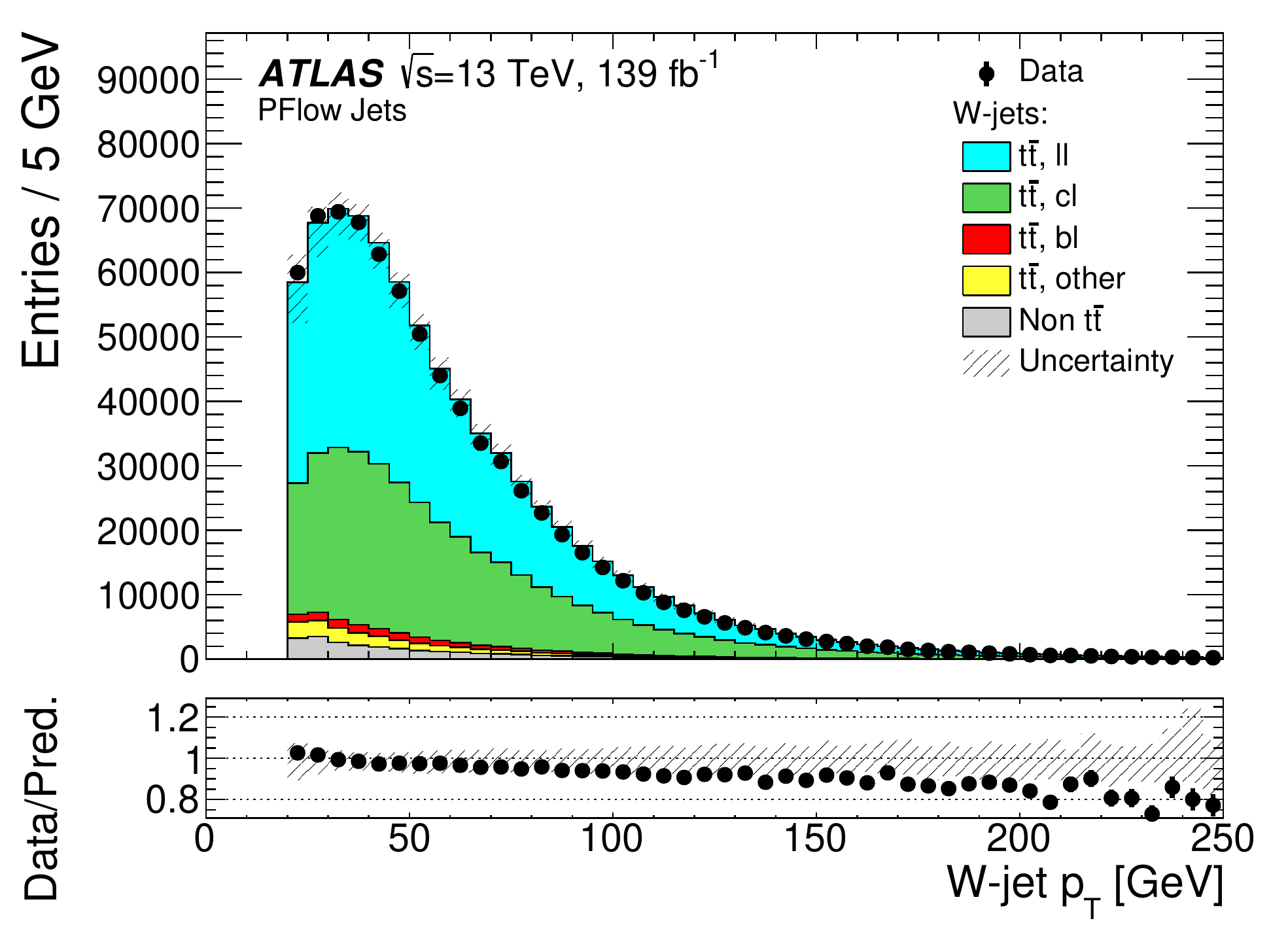}}
\subfigure[]{\includegraphics[width=0.45\textwidth]{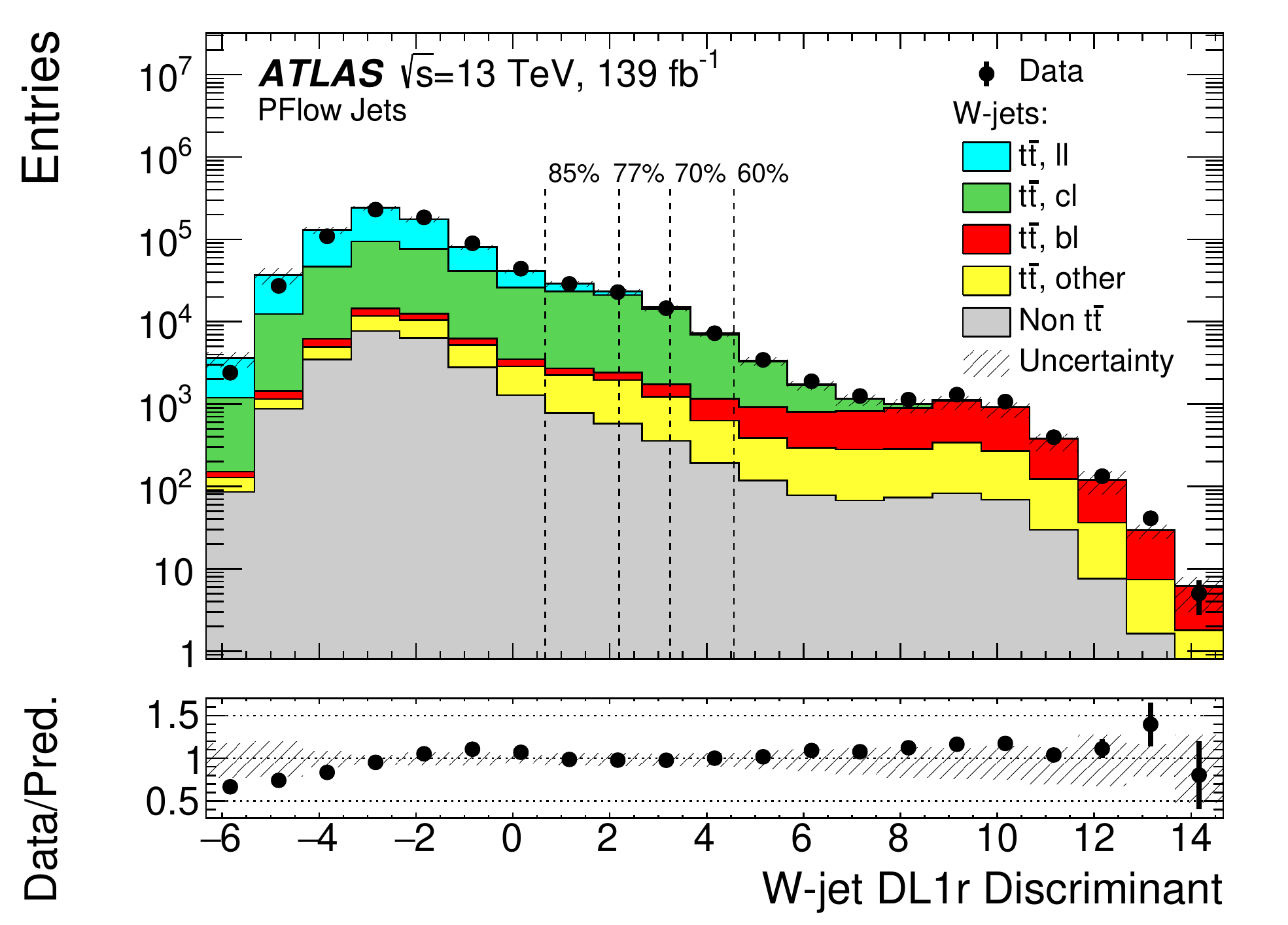}}
\caption{Distributions of  (a) $\eta$, (b) $\pt$ and (c) the DL1r discriminant
for the particle-flow jets that are associated with the $W$ boson decay
by the likelihood-based \ttbar\ event reconstruction
algorithm ($W$-jets).
The $t\bar{t}$ simulation is split according
to the flavours of the $W$-jets ($ll$, $cl$, $bl$ and other). The (mis)tagging efficiency scale factors have not been applied to the simulation. The hashed area shows the total uncertainty, excluding the uncertainties from the (mis)tagging efficiency scale factors
and the uncertainty on the jet $\pt$ distribution, which is derived from the difference between data and simulation. The vertical dashed lines in
(c) indicate the DL1r discriminant tagging intervals.}
\label{fig:plot-pfjets-1}
\end{figure}

The $\eta$ and $\pt$ distributions of the $W$-jets that are selected
with the procedure described above are shown in
Figure~\ref{fig:plot-pfjets-1}. Although the  simulation generally
describes the data well, it can be seen that it has a slightly
harder \pt\ spectrum than the data.  This difference is included as a
systematic error as described in Section~\ref{sec:systematics}.
Figure~\ref{fig:plot-pfjets-1} also shows the DL1r discriminant, where
it can be seen that the simulation describes the data reasonably well,
but there are some differences, particularly at high and low DL1r
discriminant values.
 
\section{Measurement of the charm mistagging efficiency}
\label{sec:analysis}
\subsection{Method}
The \cjet\ mistagging efficiency  is determined by
comparing the fraction of events where either of the $W$-jets are
tagged in data with the fraction tagged in simulation. The efficiency is evaluated
in four  jet-\pt\ intervals, with boundaries at 20, 40, 65, 140 and 250~\GeV,
and for the  tagging intervals
as described in Section~\ref{sec:btagging}.   Jets are
said to be tagged (untagged) if they have DL1r discriminant values
greater (less)
than the value at the 85\% boundary. The
method described below is for the `pseudo-continuous' calibration
where the efficiency is defined as the fraction of \cjets\ that have a
DL1r discriminant value that lies between the boundaries, i.e.\ a value greater than
the 60\% boundary, a value between the 70\% and 60\% boundaries, etc.
 
The light-flavour jet and \bjet\ SFs are taken from other
measurements as described in Section~\ref{sec:btagging} and the
corresponding SFs are applied to the two $W$-jets and the
two top-jets if they are light-flavour jets or \bjets. The remaining
difference between the numbers of tagged $W$-jets events in data and
simulation is used to determine the ratio of the \cjet\ mistagging efficiencies in
data and simulation.

To simplify the fit, events where
both $W$-jets are tagged, making up 0.7\% of the total number of events,
are rejected. Events where both
$W$-jets are untagged are retained to provide a normalisation of the
simulation as described below. The data are divided into bins according to
the tagging interval (labelled $t$, running from 1 to 4)  and
the $\pt$ bin of the tagged $W$-jet (labelled
$i$), and the $\pt$ bin of the untagged $W$-jet (labelled $j$).  The number of
events in each bin is denoted by $N^{t}_{\mathrm{data}}(i,j)$. In
addition, the number of untagged data events is recorded in each
$W$-jet \pt\ bin combination:
$N^{\mathrm{untag}}_{\mathrm{data}}(i,j)$ where in this case the higher-$\pt$ jet
is denoted by $j$, so $j\ge i$.
 
The fit is performed by comparing data with the total expectation, which
is dominated by $t\bar{t}$ events.
The number of simulated `signal'
events in each bin, $N^{t}_{C}(i,j)$, is made up of those events
where the tagged $W$-jet is a \cjet; the other jets in the event may be
of any flavour. There are two sources of `background'
events, defined as events where the tagged
$W$-jet is not a \cjet.  The first is the number of events where neither
of the top-jets is a \cjet, $N^{t}_{J}(i,j)$. The
second arises from events where a \cjet\ is
misclassified as one of the
top-jets. This background requires a specific treatment
since the \cjet\
is in a \pt\ bin, labelled $k$, different from those of the $W$-jets
and so must be binned in this variable in addition to $i$ and $j$. The number of
events from this background is denoted by $N^{t}_{X}(i,j,k)$.  The number of untagged simulated events is recorded in
a similar way to the data as $N^{\mathrm{untag}}_{\mathrm{MC}}(i,j)$.
 
The free parameters in the fit are the \cjet\ mistagging SF in
each \pt\ bin and tagging interval,
$c^{t}(i)$, and the overall MC normalisation in each $W$-jet
\pt\ combination, $p(i,j)$.

A fit is performed by minimising the $\chi^2$ defined as:
 
\begin{eqnarray*}
\chi^2 = \sum_{t=1}^4 \sum_{i=1}^4  \sum_{j=1}^4 \left[N^{t}_{\mathrm{data}}(i,j)- p(i,j) \Bigg ( c^{t}(i)N^{t}_{C}(i,j)+N^{t}_{J}(i,j)+\sum_{k=1}^4  c^{t=4}(k) N^{t}_{X}(i,j,k)\Bigg)\right]^2/N^{t}_{\mathrm{data}}(i,j)
\end{eqnarray*}
\begin{eqnarray}
+  \sum_{i=1}^4 \sum_{j=i}^4 \left( N^{\mathrm{untag}}_{\mathrm{data}}(i,j)-p(i,j)N^{\mathrm{untag}}_{\mathrm{MC}}(i,j)\right)^2/N^{\mathrm{untag}}_{\mathrm{data}}(i,j).
\label{eqn:chi2}
\end{eqnarray}

The background where one of the top-jets is a \cjet\   depends only on
$c^{t=4}(k)$ since the top-jets are tagged at the 60\% operating point.
The SFs for \cjets\ that are untagged in each jet \pt\ bin, $c^{\mathrm{untag}}(i)$, are evaluated  by
constraining the sum of the measured efficiencies of all tagged bins
and the untagged bin to be 1. Technically, this is evaluated by adding additional terms to the $\chi^2$ and treating the SFs
$c^{\mathrm{untag}}(i)$  as additional free parameters with
an arbitrary tolerance of $1\%$:
 
\begin{eqnarray}
\sum_{i=1}^4  \left( 1-c^{\mathrm{untag}}(i)\epsilon^\mathrm{untag}_\mathrm{MC}(i) -\sum_{t=1}^4 c^{t}(i) \epsilon_\mathrm{MC}^t(i)\right) ^2/(1\%)^2,
\label{eqn:extraterms}
\end{eqnarray}
 
where $\epsilon^\mathrm{untag}_\mathrm{MC}(i)$ and
$\epsilon_\mathrm{MC}^t(i)$ are the MC untagged and tagged
efficiencies, respectively. As part of validating the method it was
checked that scale factors of unity are obtained when performing a fit
with data replaced by the MC simulation.

\subsection{Systematic uncertainties}
\label{sec:systematics}
Several sources of experimental and MC modelling systematic
uncertainties affecting the \cjet\ efficiency SFs are
considered. The uncertainty from each source is evaluated by making
the corresponding change to the MC model and rederiving the SFs for each tagging
interval and jet \pt\ bin. The total
systematic uncertainty for each bin is  evaluated by adding the individual
contributions in quadrature.
 
The uncertainties in the jet energy scale and resolution are based on
their measurements in data~\cite{JETM-2018-05} and result in shifts of up to 5\% in the measured SFs due to a single component.
Uncertainties in the track-based soft
term of the \MET~\cite{ATLAS-CONF-2018-023} result in shifts of up to 3\%.  The
uncertainty due to the pile-up modelling~\cite{ATL-PHYS-PUB-2019-026} leads to
shifts of up to 0.7\%.  Uncertainties originating from the
electron~\cite{EGAM-2018-01} and muon~\cite{MUON-2018-03}
reconstruction lead to shifts up to 0.5\%. An additional uncertainty is applied to
cover the $W$-jet \pt\ spectrum differences between data and simulation,
giving a maximum uncertainty in the measured SFs of 0.2\%.

The uncertainties in the light-flavour mistagging and \bjet\ tagging
SFs are accounted for by shifting the SFs by each
component of the total uncertainty, and where appropriate correlating
this with the uncertainty used in the charm SF determination. The most
important non-correlated uncertainties all come from the light-flavour SF.
They originate from using the flipped tagger rather than
standard tagger, giving uncertainties of up to 7\%, and  from varying the
charm SFs in the light-flavour SF determination,
giving uncertainties of up to 4\%~\cite{ATLAS-CONF-2018-006}.

Uncertainties in the modelling of the $t\bar{t}$ background are
estimated by replacing the nominal MC sample by alternative MC
samples.  The uncertainty due to the choice of parton shower and
hadronisation model is estimated by replacing the standard {\PYTHIA}\,8
model with \herwig. This is the largest source of uncertainty
in the analysis and leads to shifts of up to 13\% in the SF.
The uncertainty in the modelling of
initial- and final-state radiation is assessed by increasing the
radiation by doubling the \hdamp\
parameter~\cite{ATL-PHYS-PUB-2016-020} and halving the renormalisation
and factorisation scales and decreasing the radiation by increasing
the scales by a factor of two.  The resulting uncertainty is at most
1\%. As already mentioned, the uncertainty in the normalisation of the
$t\bar{t}+b\bar{b}$ process~\cite{TOPQ-2017-12}
is estimated by scaling the events where both the top-jets and at
least one of the $W$-jets are \bjets\ by $\pm 25$\%, giving an uncertainty of
up to 9\% on the SF.
The uncertainty in non-$t\bar{t}$ processes is evaluated by scaling
these contributions by $\pm 25\%$, leading to a maximum uncertainty of 1\% on the SF.

\subsection{Results}
\label{sec:results}
\begin{figure}[htbp]
\centering
\includegraphics[scale = 0.4]{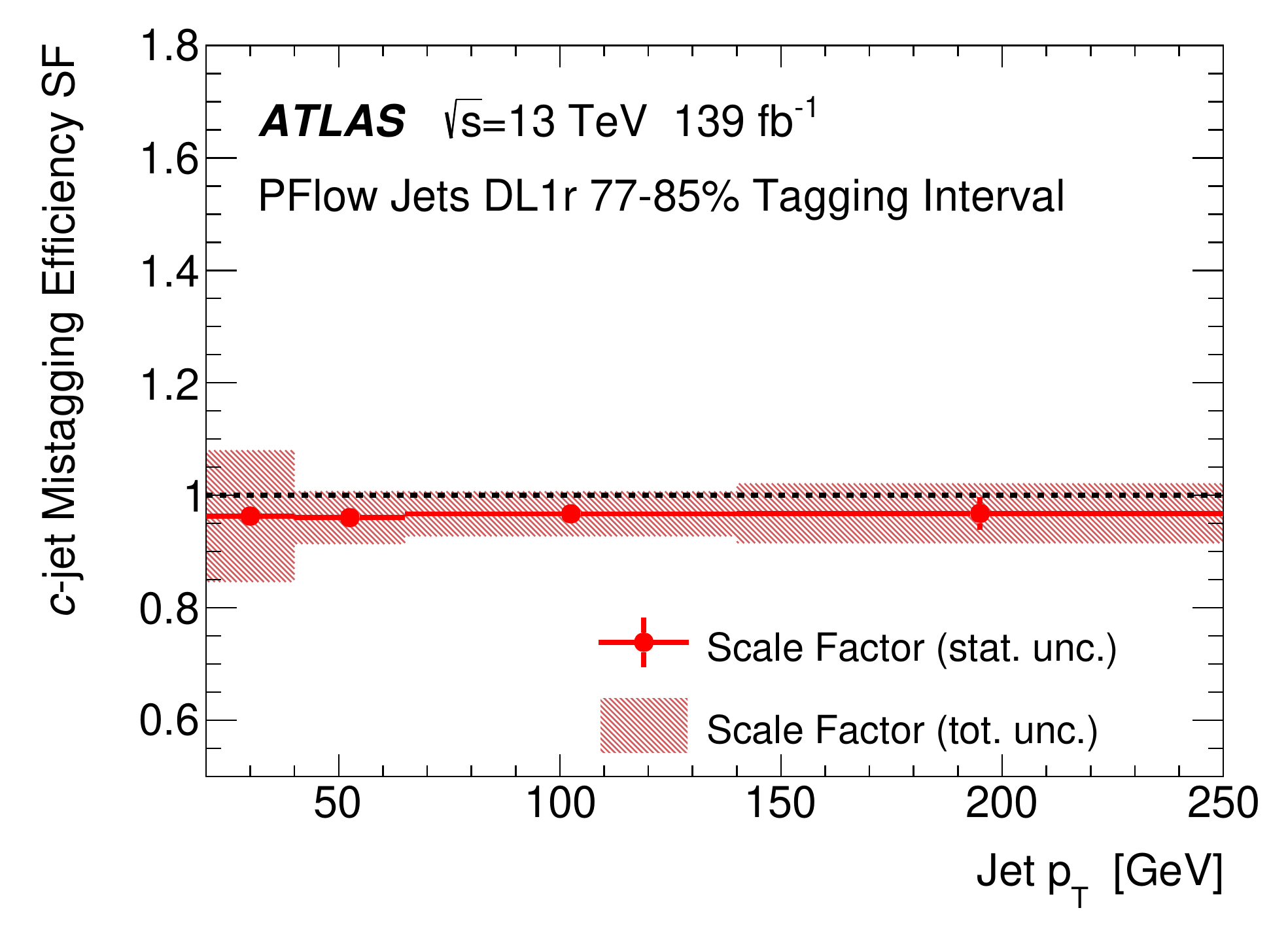}
\includegraphics[scale = 0.4]{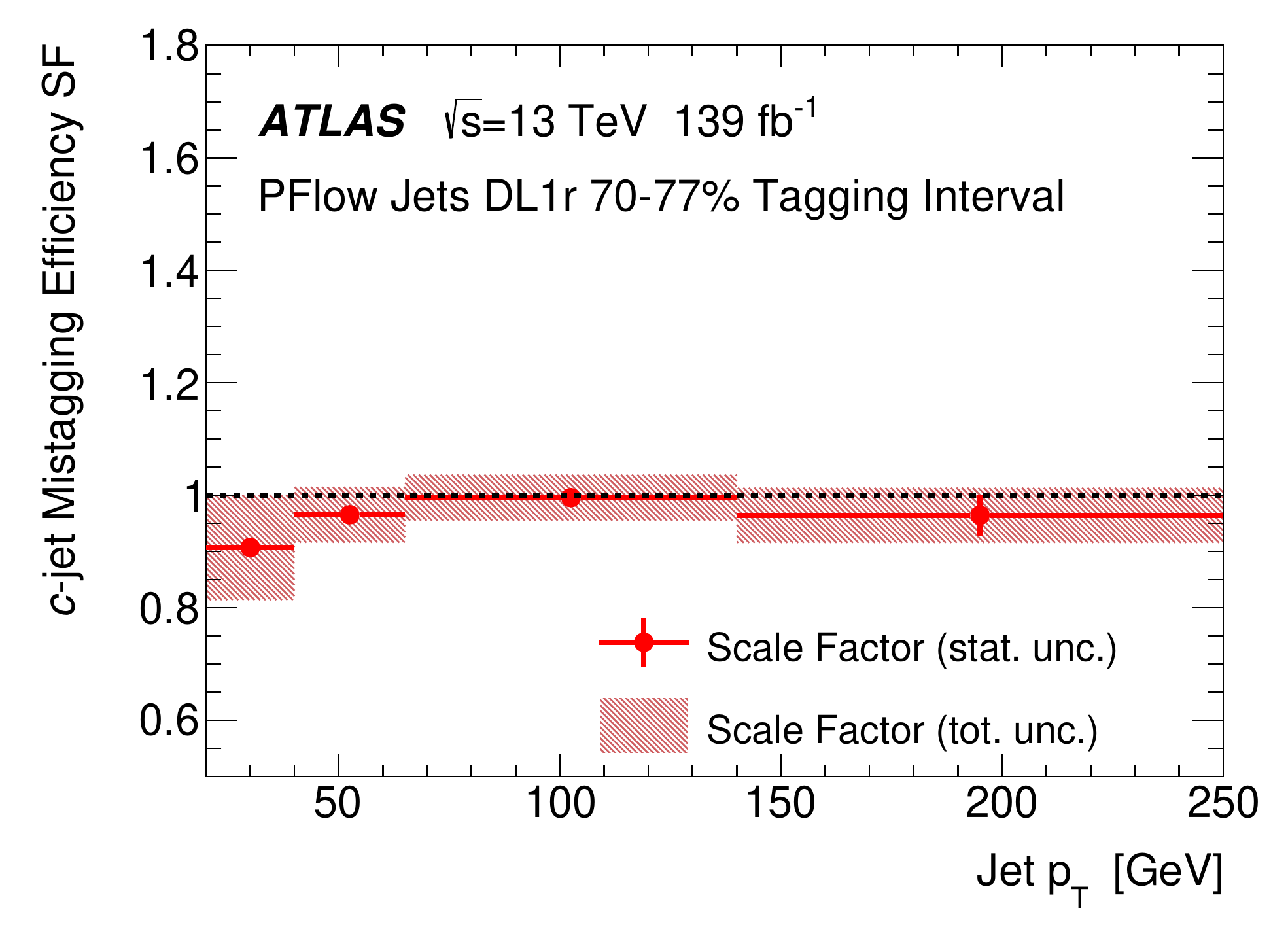}
\includegraphics[scale = 0.4]{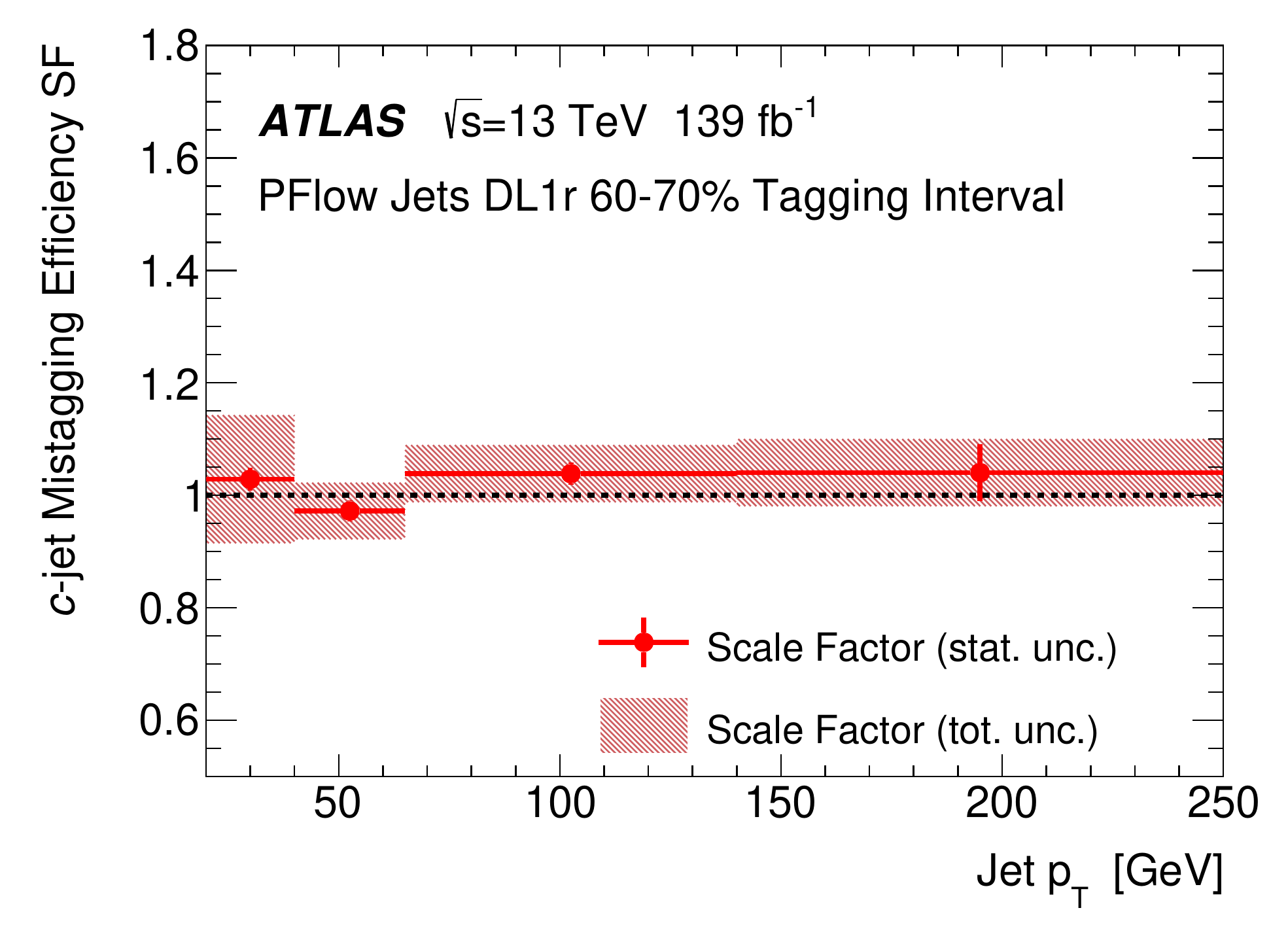}
\includegraphics[scale = 0.4]{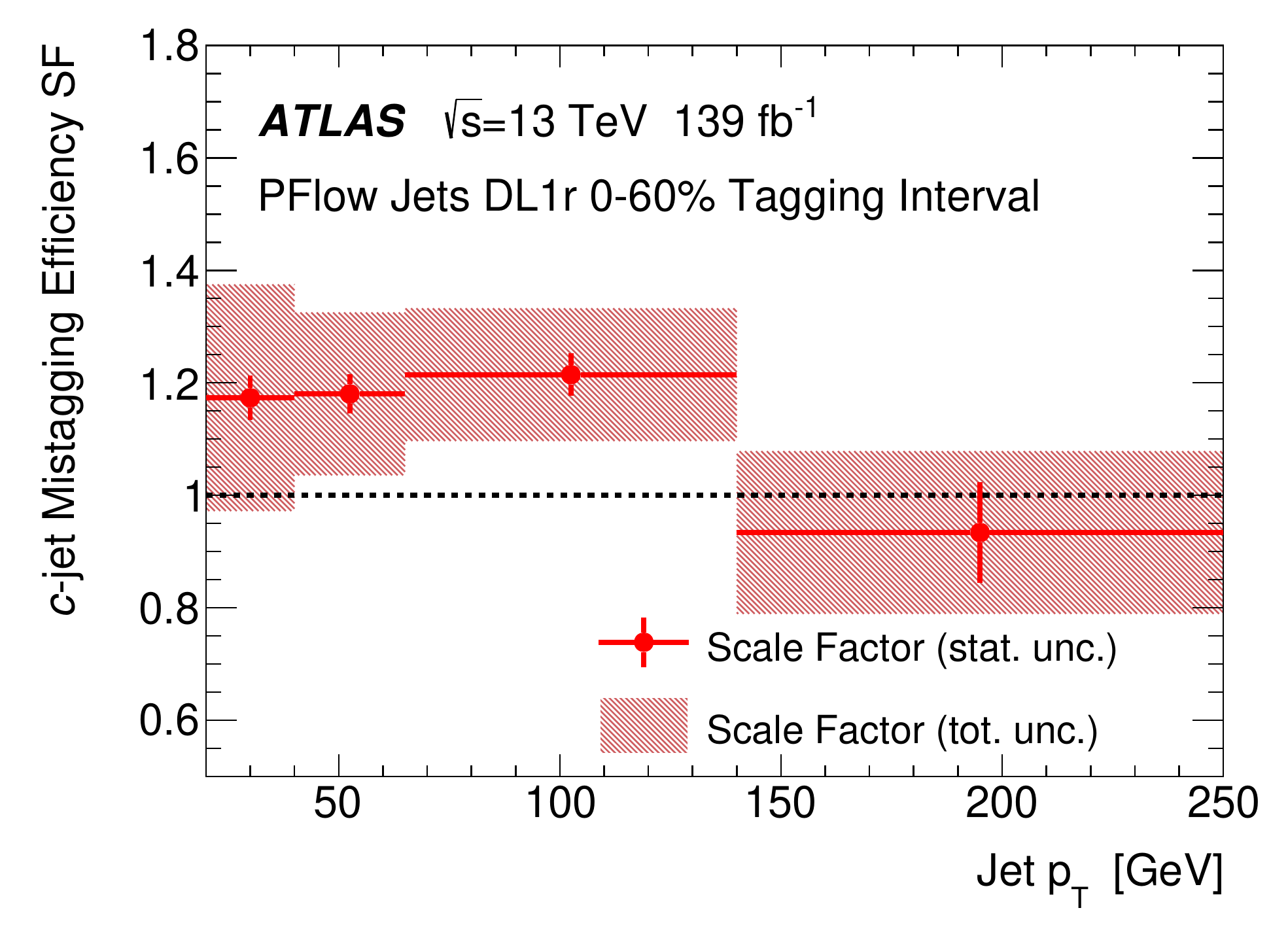}
\caption{ The \cjet\ pseudo-continuous mistagging efficiency scale factors for
particle-flow jets  shown as a function of jet \pt\ for  four
tagging intervals.  The scale factors are shown for  the {\PYTHIA}\,8+\evtgen fragmentation model.}
\label{fig:PCSFs}
\end{figure}

\begin{figure}[htbp]
\centering
\includegraphics[scale = 0.4]{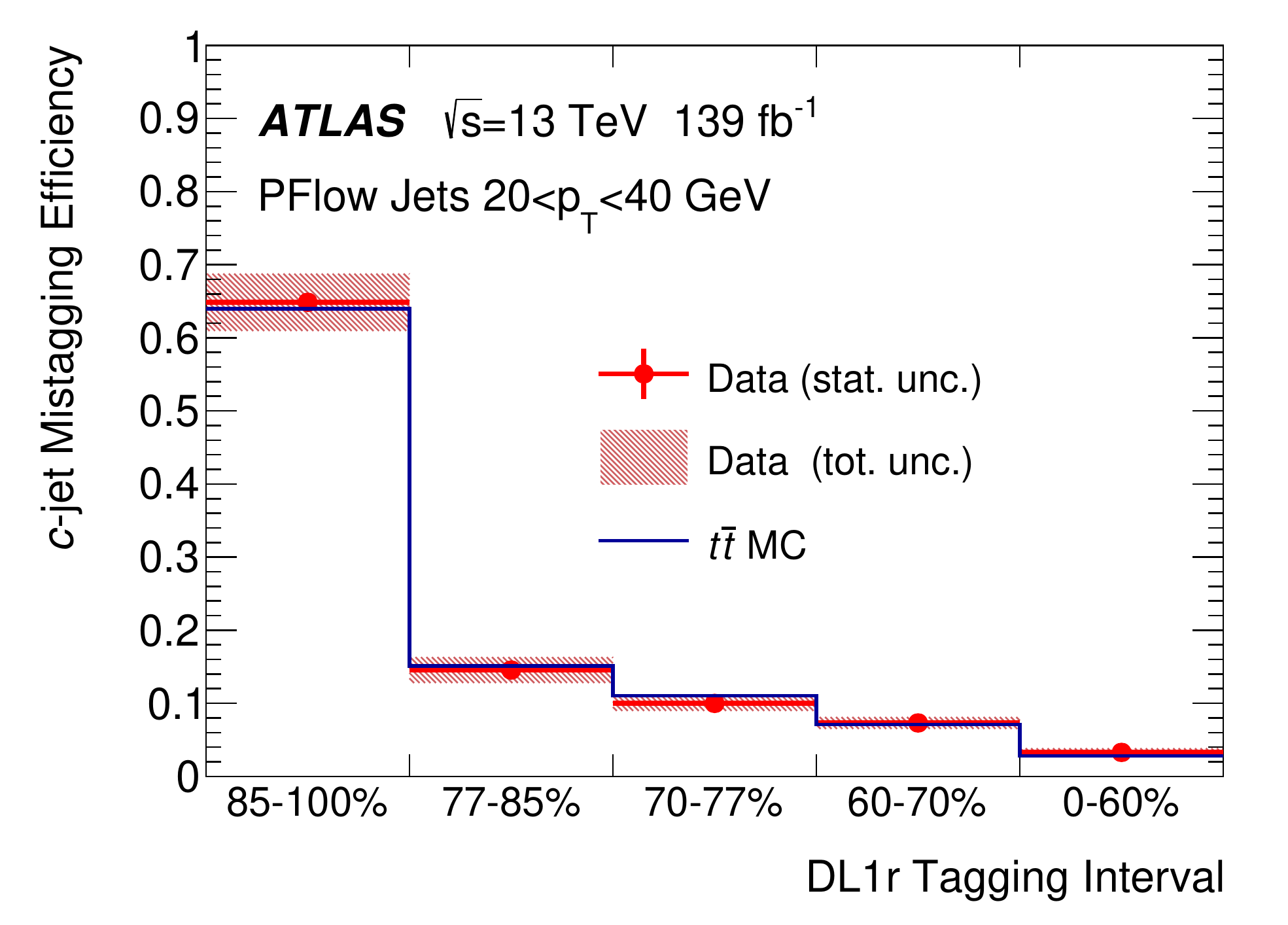}
\includegraphics[scale = 0.4]{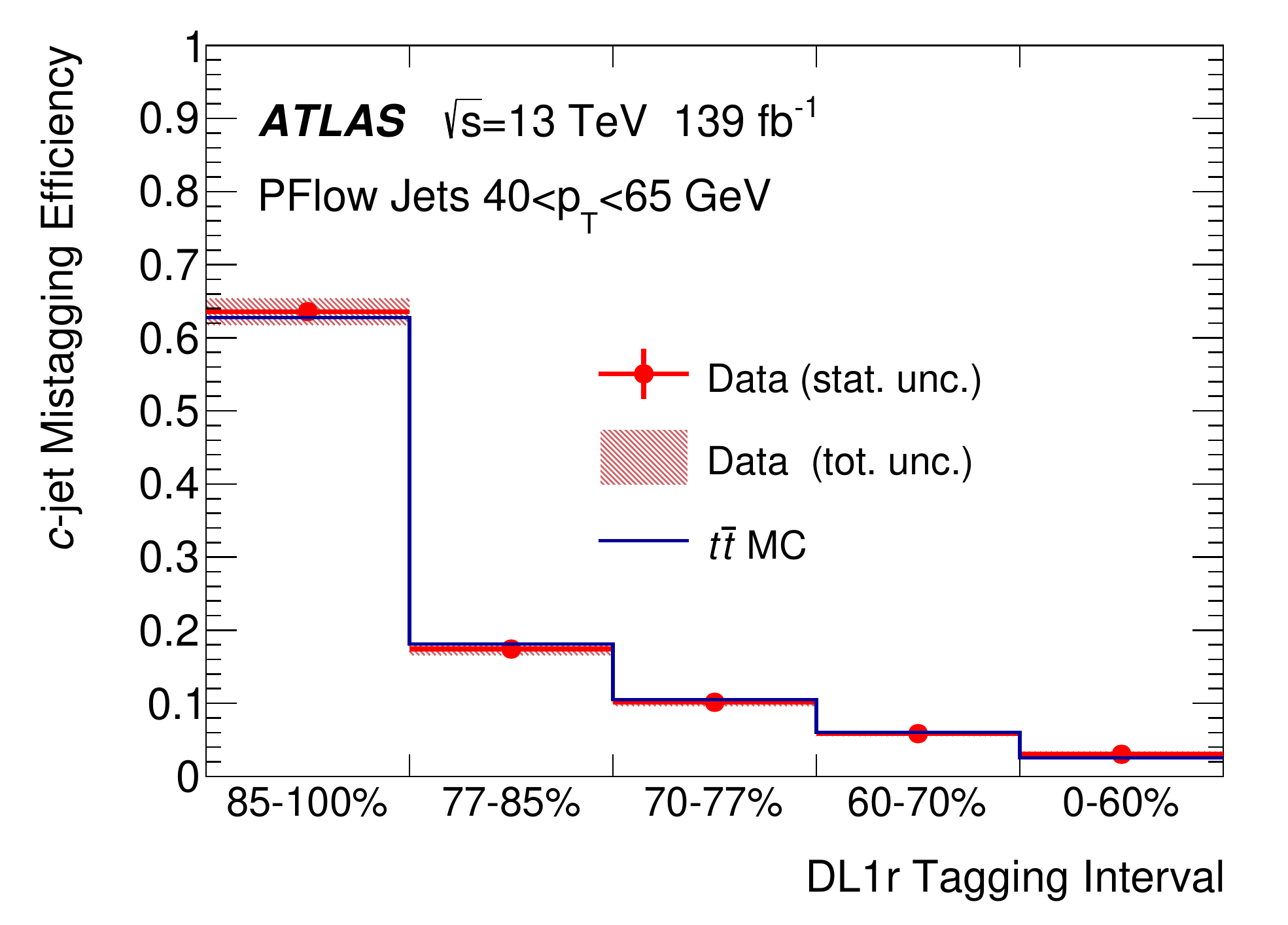}
\includegraphics[scale = 0.4]{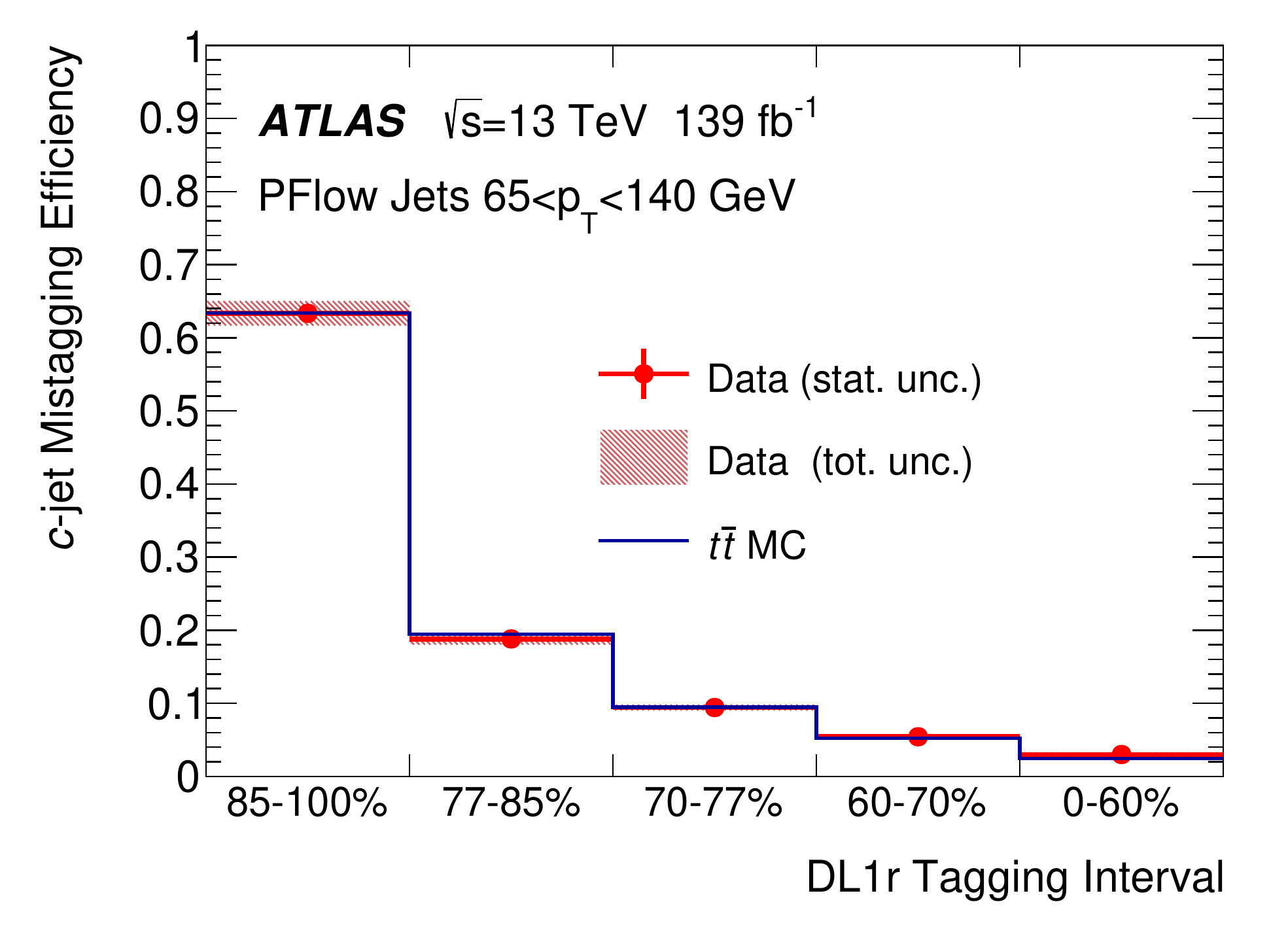}
\includegraphics[scale = 0.4]{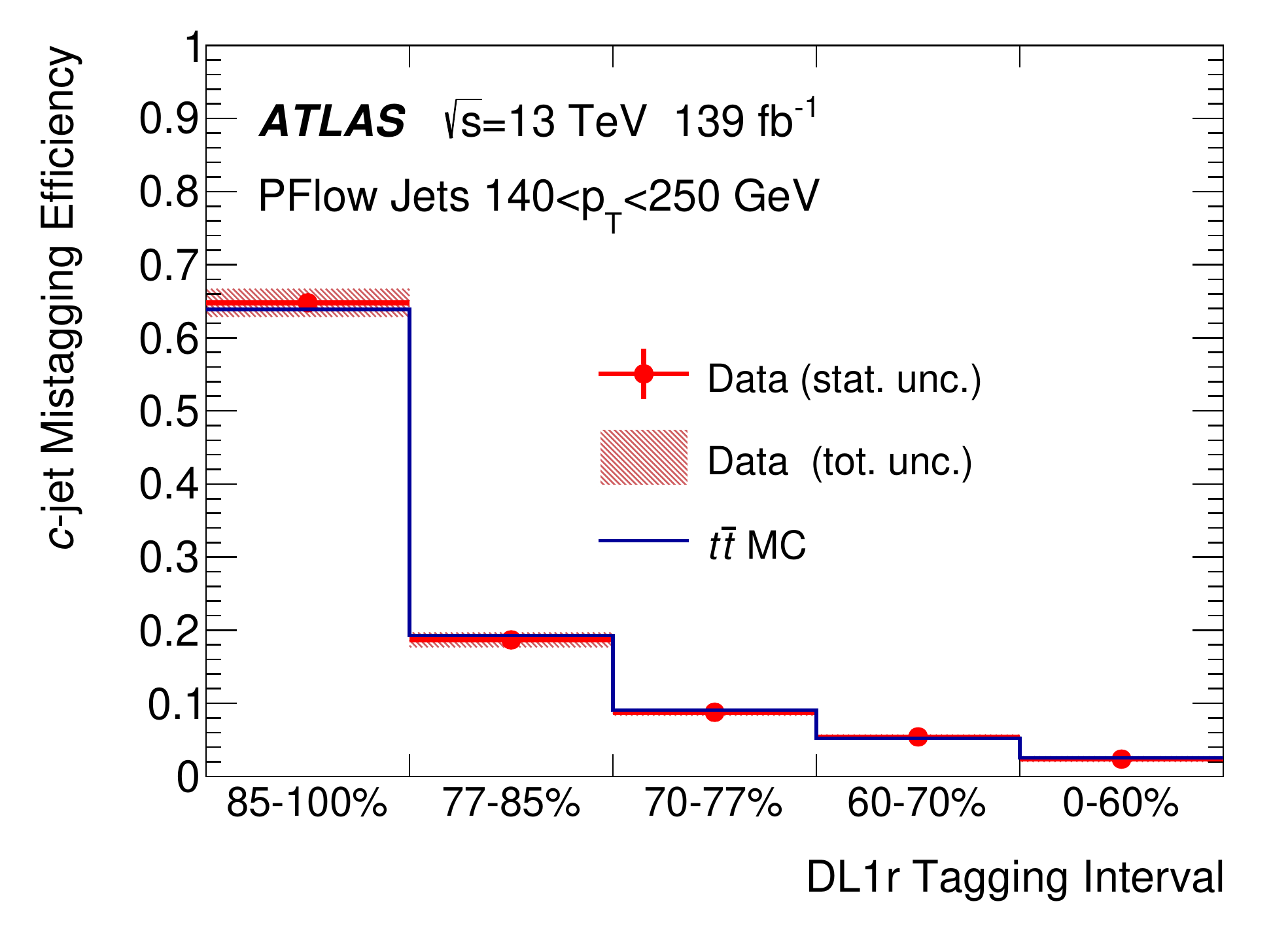}
\caption{ The \cjet\ pseudo-continuous mistagging efficiencies for
particle-flow jets shown as red points as a function of tagging interval
for the four jet-\pt\ ranges.
Included in the plots are the corresponding MC efficiencies using the
{\PYTHIA}\,8+\evtgen fragmentation model.}
\label{fig:PCeffs}
\end{figure}

The measured \cjet\ pseudo-continuous mistagging efficiency SFs
together with the total, statistical and systematic uncertainties are
listed in Table~\ref{tab:SF} and shown in Figure~\ref{fig:PCSFs} for
four tagging intervals. The fitted normalisation factors $p(i,j)$ lie in the range 0.83 to 1.05. The contributions to the systematic
uncertainty are listed in Table~\ref{tab:errors}.  The corresponding
measured data mistagging efficiencies, shown in
Figure~\ref{fig:PCeffs}, are obtained by multiplying the mistagging
efficiencies from simulation by the SFs.  The SFs do not have a strong
jet-\pt\ dependence and are consistent with unity for all but the tightest
tagging interval, where the efficiency is higher in data than in MC
simulation.  The total uncertainty varies from 3\% to 17\% and is
dominated by the systematic uncertainty except in the highest \pt\
bin.
 
\begin{table}[htbp]
\centering
\caption{The measured \cjet\ pseudo-continuous mistagging efficiency scale factors listed with the total, statistical and systematic uncertainties for particle-flow jets shown for each tagging and jet-$\pt$ interval. The scale factors are listed for  the {\PYTHIA}\,8+\evtgen fragmentation model.}
\vspace{0.2cm}
\begin{tabular}{cc|c|ccc}
Tagging interval  & Jet $\pt$ range [\GeV] & SF & Tot. & Stat. & Sys. \\ \hline
% The next lines are included from the .//tables/resultstablefull_pflow.tex input file
~~85\%--100\% & 20--40 &1.014     & 6.0\% & 1.6\% & 5.8\% \\
~~85\%--100\% & 40--65 &1.013     & 2.9\% & 1.6\% & 2.4\% \\
~~85\%--100\% & ~~65--140 &0.999  & 2.6\% & 1.6\% & 2.1\% \\
~~85\%--100\% & 140--250 &1.014   & 3.0\% & 2.1\% & 2.2\% \\   \hline
77\%--85\% & 20--40 &0.963        & 12\%~~~~~  & 1.7\% & 12\%~~~~~ \\
77\%--85\% & 40--65 &0.960        & 4.8\% & 1.1\% & 4.7\% \\
77\%--85\% & ~~65--140 &0.967     & 4.1\% & 1.1\% & 4.0\% \\
77\%--85\% & 140--250 &0.968      & 5.4\% & 2.9\% & 4.6\% \\   \hline
70\%--77\% & 20--40 &0.907        & 10\%~~~~~  & 1.6\% & 10\%~~~~~ \\
70\%--77\% & 40--65 &0.965        & 5.1\% & 1.4\% & 4.9\% \\
70\%--77\% & ~~65--140 &0.996     & 4.0\% & 1.4\% & 3.8\% \\
70\%--77\% & 140--250 &0.964      & 5.0\% & 3.7\% & 3.4\% \\   \hline
60\%--70\% & 20--40 &1.028        & 11\%~~~~~  & 2.0\% & 11\%~~~~~ \\
60\%--70\% & 40-65 &0.972         & 5.1\% & 1.8\% & 4.8\% \\
60\%--70\% & ~~65--140 &1.038     & 4.9\% & 2.0\% & 4.5\% \\
60\%--70\% & 140--250 &1.040      & 6.0\% & 5.1\% & 3.1\% \\   \hline
~~0\%--60\% & 20--40 &1.173       & 17\%~~~~~  & 3.9\% & 17\%~~~~~ \\
~~0\%--60\% & 40--65 &1.180       & 13\%~~~~~  & 3.5\% & 12\%~~~~~ \\
~~0\%--60\% & ~~65--140 &1.214    & 10\%~~~~~  & 3.8\% & 9.2\% \\
~~0\%--60\% & 140--250 &0.934     & 15\%~~~~~  & 9.0\% & 12\%~~~~~ \\
% End of text imported from the .//tables/resultstablefull_pflow.tex input file
\end{tabular}
\label{tab:SF}
\end{table}

\begin{table}[htbp]
\centering
\caption{The contributions to the
\cjet\ pseudo-continuous mistagging efficiency scale factor systematic uncertainties for
particle-flow jets. Listed are the uncertainties related to the $t\bar{t}$ modelling, the jets and \MET, the light-flavour jet scale factor, the \bjet\
scale factor, and all other sources. }
\vspace{0.2cm}
\begin{tabular}{cc|ccccc}
&   & \multicolumn{5}{c}{Systematic uncertainty in SF} \\
Tagging interval  & Jet $\pt$ range [\GeV] &  $t\bar{t}$ mod. &  Jet/\MET & Light tag & $b$-tag  & Other \\  \hline
% The next lines are included from the .//tables/systablefull_pflow.tex input file
~~85\%--100\% & 20--40 &5.3\% & 0.8\% & 2.3\% & --  & 0.1\% \\
~~85\%--100\% & 40--65 &2.1\% & 0.3\% & 1.2\% & --  & -- \\
~~85\%--100\% & ~~65--140 &1.9\% & 0.2\% & 0.9\% & --  & 0.1\% \\
~~85\%--100\% & 140--250 &2.0\% & 0.4\% & 0.9\% & --  & 0.1\% \\  \hline
77\%--85\% & 20--40 &8.3\% & 3.1\% & 8.2\% & --  & 0.3\% \\
77\%--85\% & 40--65 &3.0\% & 1.0\% & 3.5\% & --  & 0.1\% \\
77\%--85\% & ~~65--140 &3.2\% & 0.5\% & 2.4\% & --  & 0.2\% \\
77\%--85\% & 140--250 &3.9\% & 0.9\% & 2.3\% & --  & 0.3\% \\   \hline
70\%--77\% & 20--40 &9.7\% & 1.2\% & 2.5\% & --  & 0.4\% \\
70\%--77\% & 40--65 &4.7\% & 0.8\% & 1.1\% & --  & 0.1\% \\
70\%--77\% & ~~65--140 &3.7\% & 0.4\% & 0.8\% & --  & 0.2\% \\
70\%--77\% & 140--250 &3.1\% & 1.0\% & 0.9\% & --  & 0.1\% \\   \hline
60\%--70\% & 20--40 &11\%~~~~~ & 0.8\% & 1.6\% & --  & 0.4\% \\
60\%--70\% & 40--65 &4.7\% & 0.6\% & 0.9\% & --  & 0.1\% \\
60\%--70\% & ~~65--140 &4.4\% & 0.4\% & 0.6\% & --  & 0.2\% \\
60\%--70\% & 140--250 &2.6\% & 1.4\% & 0.8\% & 0.2\%  & 0.2\% \\   \hline
~~0\%--60\% & 20--40 &17\%~~~~~ & 3.1\% & 0.8\% & 0.2\%  & 1.5\% \\
~~0\%--60\% & 40--65 &11\%~~~~~ & 3.7\% & 0.4\% & 0.1\%  & 0.9\% \\
~~0\%--60\% & ~~65--140 &8.7\% & 3.0\% & 0.2\% & --  & 0.7\% \\
~~0\%--60\% & 140--250 &9.7\% & 7.1\% & 0.7\% & --  & 1.7\% \\
% End of text imported from the .//tables/systablefull_pflow.tex input file
\end{tabular}
\label{tab:errors}
\end{table}

\begin{figure}[htbp]
\centering
\includegraphics[scale = 0.4]{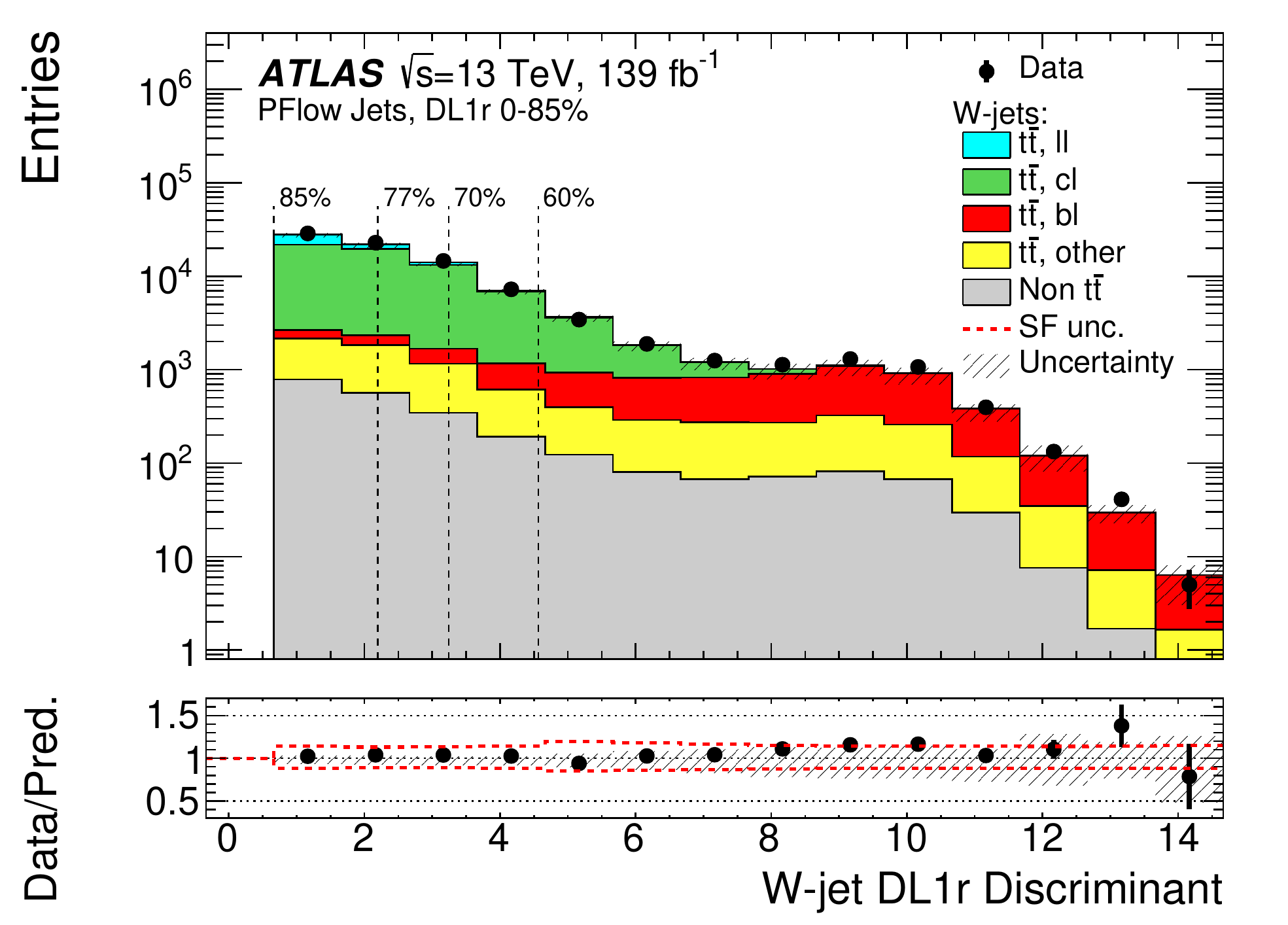}
\caption{The DL1r discriminant distribution for tagged particle-flow
jets associated with the $W$ boson decay ($W$-jets)
after applying the $b$-, $c$- and light-flavour jet tagging
scale factors to the simulation. The red dashed lines (bottom) show the
uncertainty from the jet tagging scale factors, while the hashed area shows
all other uncertainties excluding the uncertainty on the jet $\pt$ distribution, which is derived from the difference between data and simulation. The vertical dashed lines
indicate the DL1r discriminant tagging intervals.}
\label{fig:postfitdl1r}
\end{figure}

Figure~\ref{fig:postfitdl1r} shows the DL1r discriminant distributions for tagged $W$-jets,
after applying to simulation the light-flavour mistagging and $b$-tagging
efficiency SFs, taken from other measurements as
described in Section~\ref{sec:btagging}, and the \cjet\ SFs obtained
from this analysis. It can be seen that
the simulation agrees with the data within uncertainties over the entire distribution.
 
\begin{figure}[htbp]
\centering
\includegraphics[scale = 0.4]{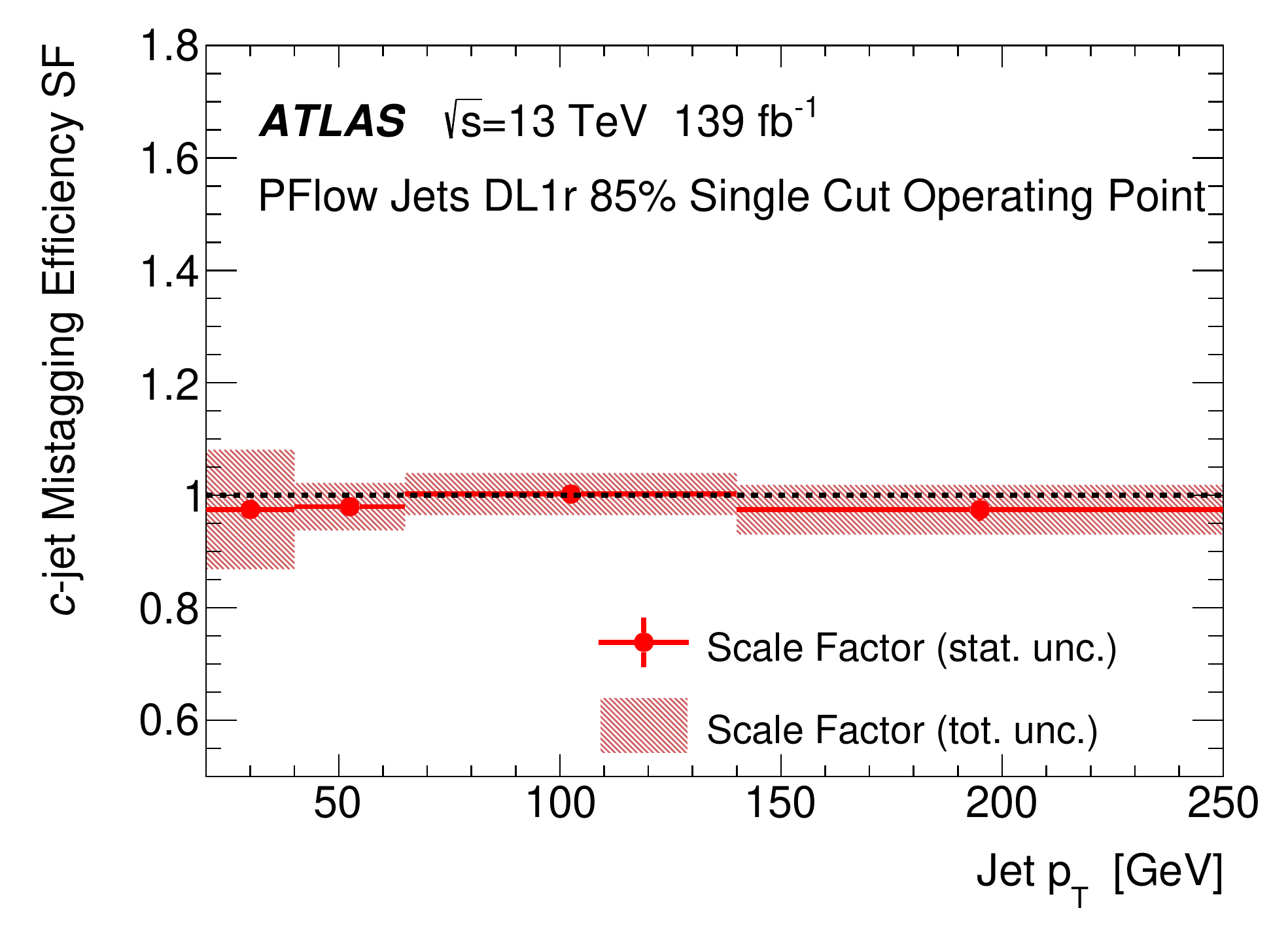}
\includegraphics[scale = 0.4]{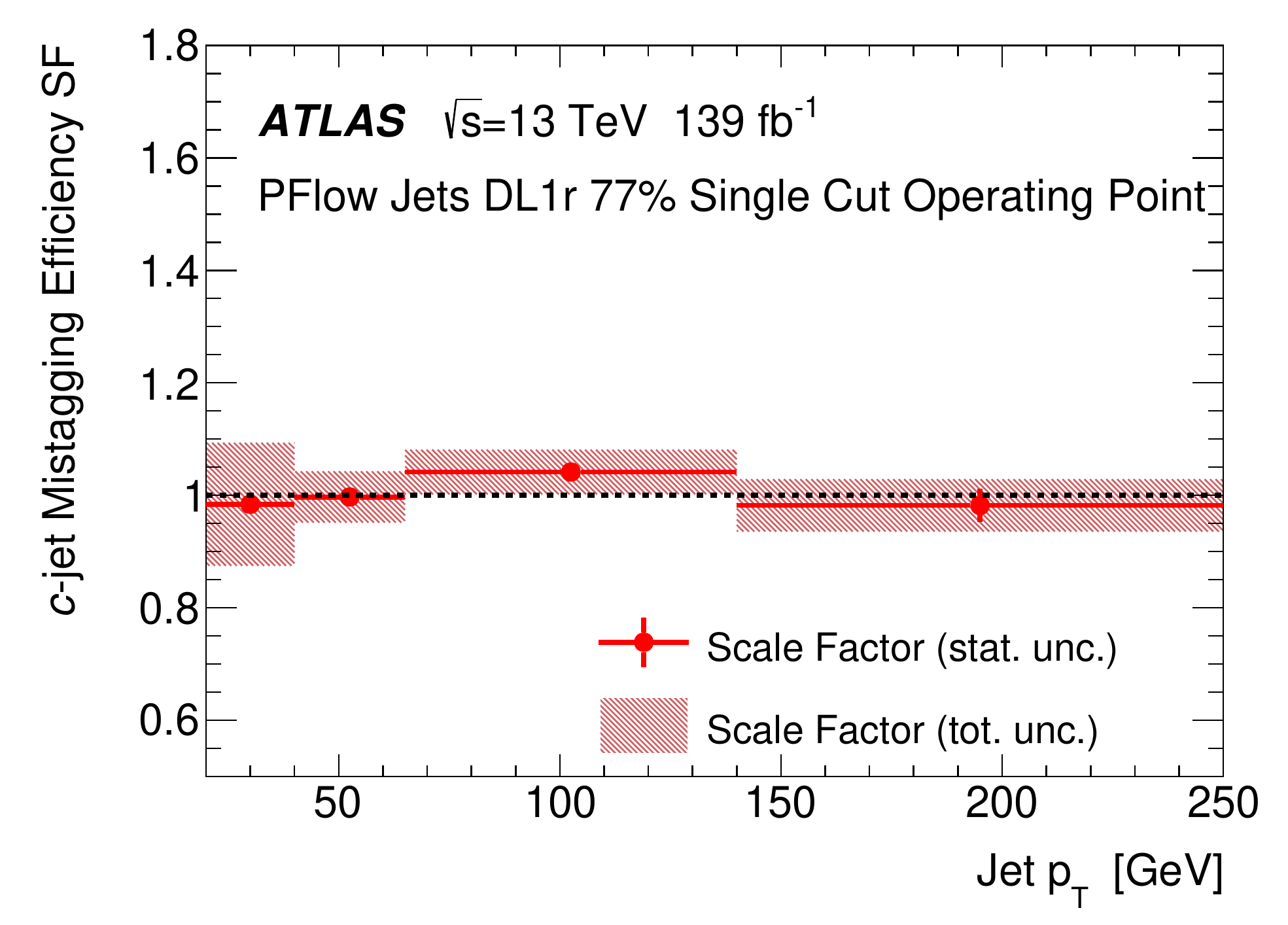}
\includegraphics[scale = 0.4]{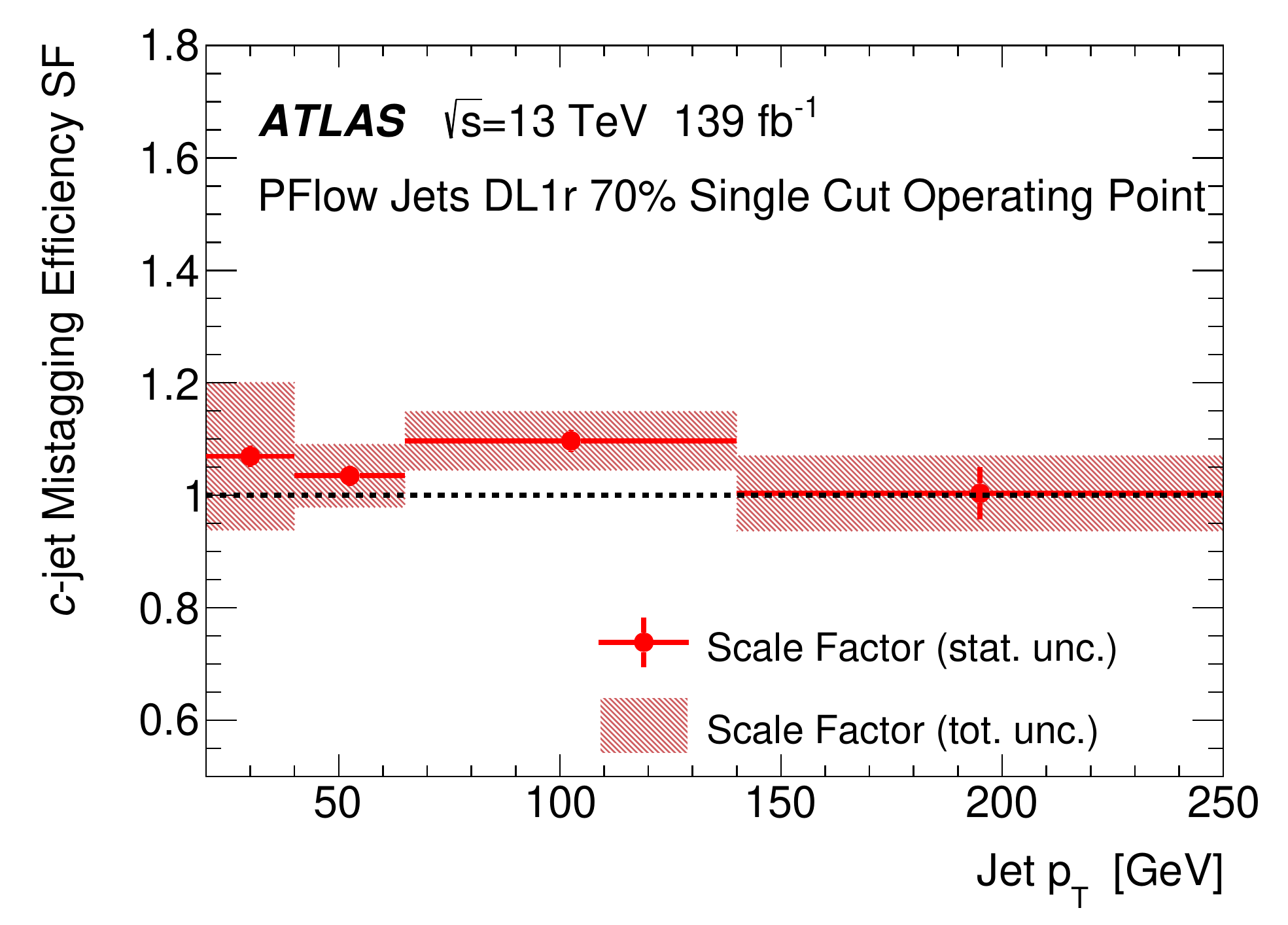}
\includegraphics[scale = 0.4]{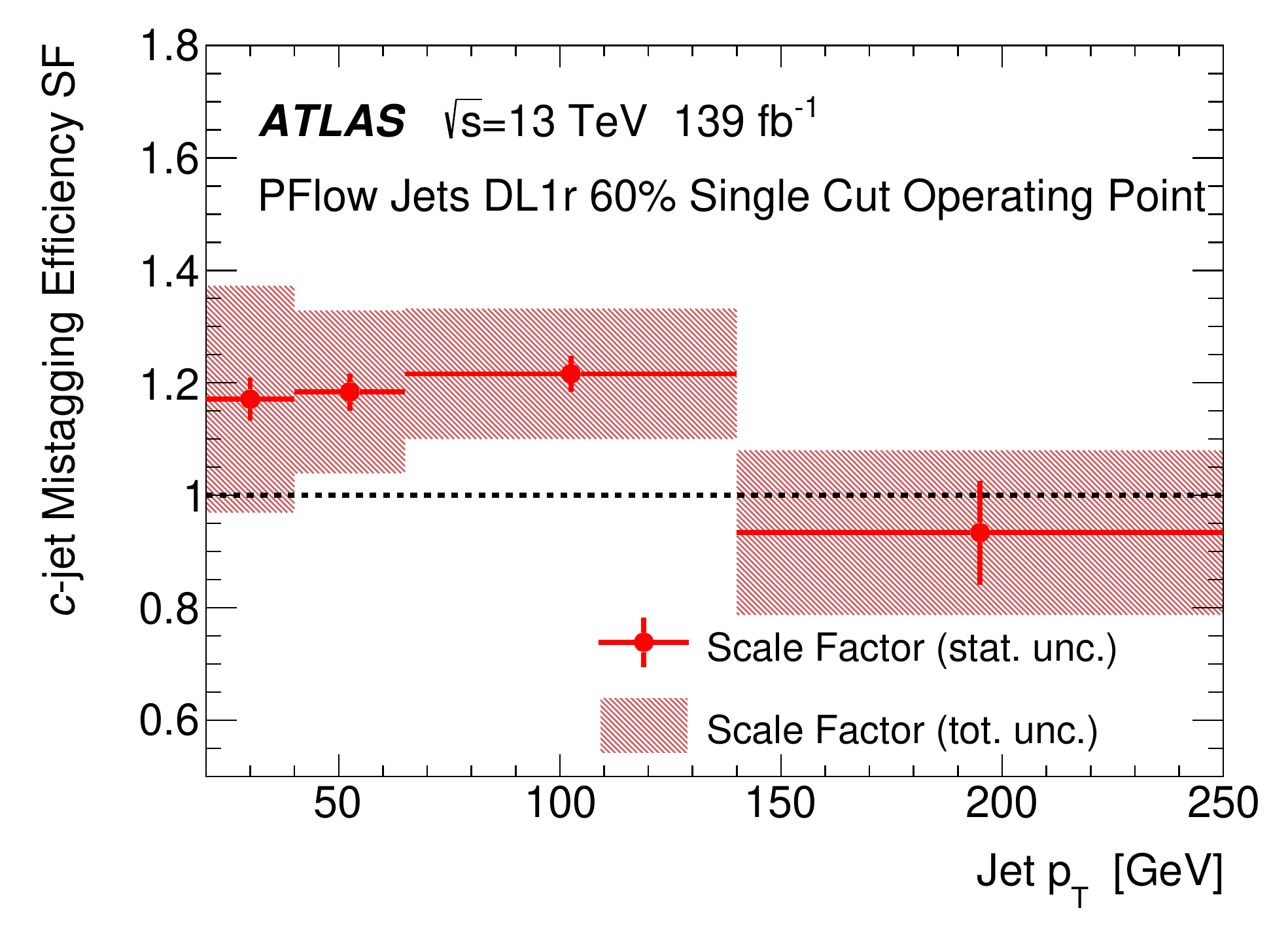}
\caption{ The \cjet\ single-cut mistagging efficiency scale factors for
particle-flow jets
shown as a function of jet \pt\ for the four
tagging operating points.  The scale factors are shown for  the {\PYTHIA}\,8+\evtgen fragmentation model.}
\label{fig:SFsnonPC}
\end{figure}

Many ATLAS analyses also use the SFs in a simplified form
called `single cut', where events are classified as
tagged (untagged) if the DL1r  discriminant lies above (below) one
of the operating point boundaries. The corresponding
SFs are determined in
a similar way to the \cjet\ pseudo-continuous mistagging efficiency SFs by using Eq.~(\ref{eqn:chi2}),
but replacing $N^{t}_{\mathrm{data}}$, $N^{t}_{C}$, $N^{t}_{J}$ and
$N^{t}_{X}$ with the corresponding numbers of events in which the
tagged jet has a DL1r discriminant value above the operating point. The extra terms
in Eq.~(\ref{eqn:extraterms}) are not used in these fits. Instead the untagged
SFs are evaluated using the results of the fit as $(1-c^t(i)\epsilon^t_\mathrm{MC}(i))/(1-\epsilon^t_\mathrm{MC}(i))$.
The resulting
single-cut \cjet\ mistagging efficiency SFs are shown in
Figure~\ref{fig:SFsnonPC}.

\FloatBarrier
\section{Conclusion}
\label{sec:conclusion}
A new technique has been used to measure the inclusive \cjet\ mistagging efficiency,
using semileptonic $t\bar{t}$ events from \(pp\) collisions at $\sqrt{s}=13$~\TeV\ at the
LHC. The measurement is based on an integrated luminosity of 139~fb$^{-1}$, corresponding to
the full Run~2 dataset, collected with the ATLAS detector.
The efficiencies were measured as a
function of jet \pt\ in the range 20-250~GeV, with
boundaries corresponding to 60\%, 70\%, 77\% and 85\% efficiency to tag \bjets\
in simulated \ttbar events.  The
total uncertainties are in the range 3\%--17\%.
The measured efficiencies generally agree with those in
simulation but are higher for the tightest (0--60\%) tagging interval.

\section*{Acknowledgements}
 
% The next lines are included from the .//acknowledgements/Acknowledgements.tex input file

We thank CERN for the very successful operation of the LHC, as well as the
support staff from our institutions without whom ATLAS could not be
operated efficiently.
 
We acknowledge the support of
ANPCyT, Argentina;
YerPhI, Armenia;
ARC, Australia;
BMWFW and FWF, Austria;
ANAS, Azerbaijan;
SSTC, Belarus;
CNPq and FAPESP, Brazil;
NSERC, NRC and CFI, Canada;
CERN;
ANID, Chile;
CAS, MOST and NSFC, China;
Minciencias, Colombia;
MSMT CR, MPO CR and VSC CR, Czech Republic;
DNRF and DNSRC, Denmark;
IN2P3-CNRS and CEA-DRF/IRFU, France;
SRNSFG, Georgia;
BMBF, HGF and MPG, Germany;
GSRI, Greece;
RGC and Hong Kong SAR, China;
ISF and Benoziyo Center, Israel;
INFN, Italy;
MEXT and JSPS, Japan;
CNRST, Morocco;
NWO, Netherlands;
RCN, Norway;
MNiSW and NCN, Poland;
FCT, Portugal;
MNE/IFA, Romania;
JINR;
MES of Russia and NRC KI, Russian Federation;
MESTD, Serbia;
MSSR, Slovakia;
ARRS and MIZ\v{S}, Slovenia;
DSI/NRF, South Africa;
MICINN, Spain;
SRC and Wallenberg Foundation, Sweden;
SERI, SNSF and Cantons of Bern and Geneva, Switzerland;
MOST, Taiwan;
TAEK, Turkey;
STFC, United Kingdom;
DOE and NSF, United States of America.
In addition, individual groups and members have received support from
BCKDF, CANARIE, Compute Canada and CRC, Canada;
COST, ERC, ERDF, Horizon 2020 and Marie Sk{\l}odowska-Curie Actions, European Union;
Investissements d'Avenir Labex, Investissements d'Avenir Idex and ANR, France;
DFG and AvH Foundation, Germany;
Herakleitos, Thales and Aristeia programmes co-financed by EU-ESF and the Greek NSRF, Greece;
BSF-NSF and GIF, Israel;
Norwegian Financial Mechanism 2014-2021, Norway;
La Caixa Banking Foundation, CERCA Programme Generalitat de Catalunya and PROMETEO and GenT Programmes Generalitat Valenciana, Spain;
G\"{o}ran Gustafssons Stiftelse, Sweden;
The Royal Society and Leverhulme Trust, United Kingdom.
 
The crucial computing support from all WLCG partners is acknowledged gratefully, in particular from CERN, the ATLAS Tier-1 facilities at TRIUMF (Canada), NDGF (Denmark, Norway, Sweden), CC-IN2P3 (France), KIT/GridKA (Germany), INFN-CNAF (Italy), NL-T1 (Netherlands), PIC (Spain), ASGC (Taiwan), RAL (UK) and BNL (USA), the Tier-2 facilities worldwide and large non-WLCG resource providers. Major contributors of computing resources are listed in Ref.~\cite{ATL-SOFT-PUB-2021-003}.
 
% End of text imported from the .//acknowledgements/Acknowledgements.tex input file

\clearpage

\printbibliography
 
\clearpage % ATLAS Collaboration author list
% Reference date of FTAG-2020-08 is 2021-07-15
% Author list last updated on date 03-FEB-22
% Data extracted on 03-Feb-2022 for paper reference FTAG-2020-08
% at 12:10pm
 
\begin{flushleft}
\hypersetup{urlcolor=black}
{\Large The ATLAS Collaboration}

\bigskip

\AtlasOrcid[0000-0002-6665-4934]{G.~Aad}$^\textrm{\scriptsize 98}$,    
\AtlasOrcid[0000-0002-5888-2734]{B.~Abbott}$^\textrm{\scriptsize 124}$,    
\AtlasOrcid[0000-0002-7248-3203]{D.C.~Abbott}$^\textrm{\scriptsize 99}$,    
\AtlasOrcid[0000-0002-2788-3822]{A.~Abed~Abud}$^\textrm{\scriptsize 34}$,    
\AtlasOrcid[0000-0002-1002-1652]{K.~Abeling}$^\textrm{\scriptsize 51}$,    
\AtlasOrcid[0000-0002-2987-4006]{D.K.~Abhayasinghe}$^\textrm{\scriptsize 91}$,    
\AtlasOrcid[0000-0002-8496-9294]{S.H.~Abidi}$^\textrm{\scriptsize 27}$,    
\AtlasOrcid[0000-0002-9987-2292]{A.~Aboulhorma}$^\textrm{\scriptsize 33e}$,    
\AtlasOrcid[0000-0001-5329-6640]{H.~Abramowicz}$^\textrm{\scriptsize 157}$,    
\AtlasOrcid[0000-0002-1599-2896]{H.~Abreu}$^\textrm{\scriptsize 156}$
% ...
\bigskip
\\

% ...
$^{27}$Physics Department, Brookhaven National Laboratory, Upton NY; United States of America.\\
% ...
$^{33}$$^{(a)}$Facult\'e des Sciences Ain Chock, R\'eseau Universitaire de Physique des Hautes Energies - Universit\'e Hassan II, Casablanca;$^{(b)}$Facult\'{e} des Sciences, Universit\'{e} Ibn-Tofail, K\'{e}nitra;$^{(c)}$Facult\'e des Sciences Semlalia, Universit\'e Cadi Ayyad, LPHEA-Marrakech;$^{(d)}$LPMR, Facult\'e des Sciences, Universit\'e Mohamed Premier, Oujda;$^{(e)}$Facult\'e des sciences, Universit\'e Mohammed V, Rabat;$^{(f)}$Mohammed VI Polytechnic University, Ben Guerir; Morocco.\\
$^{34}$CERN, Geneva; Switzerland.\\
% ...
$^{51}$II. Physikalisches Institut, Georg-August-Universit\"{a}t G\"ottingen, G\"ottingen; Germany.\\
% ...
$^{91}$Department of Physics, Royal Holloway University of London, Egham; United Kingdom.\\
% ...
$^{98}$CPPM, Aix-Marseille Universit\'e, CNRS/IN2P3, Marseille; France.\\
$^{99}$Department of Physics, University of Massachusetts, Amherst MA; United States of America.\\
% ...
$^{124}$Homer L. Dodge Department of Physics and Astronomy, University of Oklahoma, Norman OK; United States of America.\\
% ...
$^{156}$Department of Physics, Technion, Israel Institute of Technology, Haifa; Israel.\\
$^{157}$Raymond and Beverly Sackler School of Physics and Astronomy, Tel Aviv University, Tel Aviv; Israel.\\
% ...
$^{a}$ Also at Borough of Manhattan Community College, City University of New York, New York NY; United States of America.\\
$^{b}$ Also at Bruno Kessler Foundation, Trento; Italy.\\
$^{c}$ Also at Center for High Energy Physics, Peking University; China.\\
$^{d}$ Also at Centro Studi e Ricerche Enrico Fermi; Italy.\\
$^{e}$ Also at CERN, Geneva; Switzerland.\\
$^{f}$ Also at D\'epartement de Physique Nucl\'eaire et Corpusculaire, Universit\'e de Gen\`eve, Gen\`eve; Switzerland.\\
% ...

\end{flushleft}

% Created with Glance <Atlas.Glance@cern.ch>

\clearpage

\end{document}